# The use of satellite and ground based measurements for estimating and reducing uncertainties in the spatial distribution of emissions of nitrogen oxides

I. B. Konovalov[1,2], M. Beekmann[2], A. Richter[3], J. P. Burrows[3]

[1]{Institute of Applied Physics, Russian Academy of Sciences, Nizhniy Novgorod, Russia}
[2]{ Laboratoire Inter-Universitaire de Systèmes Atmosphériques, CNRS, Créteil, France}
[3]{Institute of Environmental Physics and Remote Sensing, IUP/IFE, University of Bremen, Bremen, Germany}
Correspondence to: I. B. Konovalov (konov@appl.sci-nnov.ru)

## Abstract

We explore possibilities of improving the spatial structure of $NO_x$ emissions employed in a continental scale chemistry transport model (CTM) by using satellite measurements of nitrogen dioxide and ground-based observations of near surface ozone. In this study, we combine the tropospheric $NO_2$ columns derived from SCIAMACHY measurements, the data from the EMEP ozone-monitoring network and the calculations performed with the CHIMERE CTM in the framework of an advanced inverse modelling scheme. All data used in the study correspond to the period of June-August 2003. The main distinctive feature of our inversion scheme is that, in contrast to more common inverse modelling approaches, the magnitudes of uncertainties in the input data are not explicitly predefined but rather estimated consistently with the "a posteriori" emissions as a result of the inversion. While the tropospheric $NO_2$ columns are used for "fitting" the spatial distribution of the emission parameters of the model, the ozone observations are only used to estimate the averaged levels of uncertainties in "a priori" emissions. We use our method in order to estimate and to reduce uncertainties in the gridded (with the resolution of 1 degree) $NO_x$ emission data for Europe, Middle East and Northern Africa. It is found that the "a priori" emission estimates used in the standard version of CHIMERE are probably biased in several regions. In particular, the total emissions from both anthropogenic and biogenic sources are probably, overestimated for Great Britain, Belgium, the Netherlands, Lithuania, Latvia, Poland, Bulgaria, Greece, and Iraq, but mostly underestimated for Spain, Italy, Switzerland, Czechia, Former Yugoslavia, Turkey, Lebanon and Iran. The emission correction factors are typically inside of the range from 0.5 to 2. On average, the uncertainties in total $NO_x$ emissions are estimated to be about 1.7 in terms of the geometric standard deviation in Europe and about 2.1 outside of Europe. The corrected emission estimates provide better agreement of the modelled results with observations for both $NO_2$ columns and near surface concentrations of ozone.

## 1   Introduction

It is widely recognized that the ability of atmospheric models to represent the current state and to predict possible future changes of the chemical composition of the atmosphere depends crucially on the quality of available information regarding sources (emissions) of atmospheric pollutants. One promising way for validation of available emission inventories and for elaboration of improved emission data involves inverse modeling of sources of atmospheric gases (see, e.g., Enting, 2002). In this way, observational data are used as constraints to emission parameters of an atmospheric model.

 A strong impetus to the atmospheric inverse modeling studies has been given by recent remarkable progress in satellite measurements of the composition of the lower atmosphere. Indeed, it seems reasonable to believe that data for tropospheric column amounts of such important trace gases as $NO_2$, $SO_2$, $CH_4$, CO, HCHO derived from almost global satellite measurements (e.g., Burrows et al. 1999; Eisinger and Burrows, 1998; Velders et al. 2001; Richter and Burrows, 2002; Buchwitz et al., 2004; Bowman, 2006) bear significant useful information on sources of these gases. Attempts to extract and to use this information have already been undertaken in several studies. In particular, tropospheric $NO_2$ columns derived from GOME measurements have been used for estimation of total $NO_x$ emissions on the global scale (Leue et al., 2001; Martin et al., 2003; Müller and Stavrakou, 2005) and on the regional scale (Konovalov et al., 2006), as well as for estimations of $NO_x$ emissions from different kinds of sources such as lightning (Boersma et al. 2005; Beirle et al., 2006), ships (Beirle et al., 2004, Richter et al., 2004) or soils (Jaegle et al., 2004, Bertram et al., 2005). Satellite measurements have also been used to improve emission estimates for carbon monoxide (Pétron et al., 2004, Yurganov et al., 2005) and isoprene (Visconti et al., 2005).
In this study, we use satellite measurements performed by the SCIAMACHY satellite instrument (Bovensmann et al., 1999) in order to estimate and reduce uncertainties in available data for $NO_x$ emissions. More specifically, we consider spatially distributed $NO_x$ emission data employed in the CHIMERE continental scale air quality model covering Europe and the Middle East. A continental scale model allows us to resolve finer spatial scales than those usually considered in global inverse modelling studies. The combination of satellite data with an air quality model and data of ground-based air pollution monitoring in this study is expected to contribute to developing synergy between regional air pollution studies and observations of the chemical composition of the atmosphere from space.

A main idea underlying atmospheric inverse modelling studies is to adjust emission parameters of a model such that model results would demonstrate closer agreement with observations. However, in doing so it is necessary to keep in mind that both the model and measurements are imperfect. Accordingly, it is crucial to insure that fitting of emissions does not result in compensation of model and measurement biases that are actually not related to uncertainties in emission data. Otherwise, the fitted emissions may even be more uncertain than available emissions estimates. A common way to avoid this undesirable scenario is to use a probabilistic approach that usually involves a combination of "a priori" information on the considered emissions and information derived from measurements using a model (e.g., Enting, 2002). The "weights" of these sources of information relate to uncertainties in the a priori estimates, model and measurement errors. Although these weights are recognised to be crucial parameters of inverse modelling schemes (Kaminski, 1999; Heimann and Kaminski, 1999), they are usually assigned in a rather arbitrary way as a kind of "expert estimates". However, neither model errors, nor uncertainties in available emission inventories are ever known sufficiently well. It is useful to note, in particular, that while overall uncertainties in the model can be judged by considering differences between model results and measurements, it is necessary to take into account that these differences actually include both model errors caused by uncertainties in emissions and all other model errors. Different parts of the model errors may, in principle, covariate or anti-covariate making separate estimation of these parts of uncertainties difficult or, perhaps, even impossible. The uncertainties in satellite data concerning the chemical composition of the lower atmosphere are also difficult to quantify, since they result from a rather complex retrieval procedure that involves important a priori assumptions regarding vertical profiles of the species concentration, aerosol and cloud properties. Although the attempts to estimate uncertainties in satellite data or emission inventories using the error propagation technique (see, e.g., Boersma et al, 2004; Kühlwein et al., 2000) are undoubtedly useful, these studies also involve explicit authors' assumptions about the magnitudes of the uncertainties in input data used either in their retrieval procedure or in the respective emission inventory.

It has been suggested recently (Konovalov et al., 2006) that the uncertainties mentioned above can be estimated consistently within the inverse modeling scheme. In other words, it has been proposed that the parameters of the corresponding probability distribution functions can be treated as internal rather than external parameters of the inversion scheme. To implement this idea, the inverse modeling scheme involved independent measurements of different characteristics (ground based in addition to satellite measurements); that allowed deducing optimal values of uncertainties in model results, measurements and a priori emissions. In the given study, the previously proposed methods are substantially advanced. The major distinctive features of this study are (i) the use of ground-based measurements of near-surface ozone concentrations (instead of $NO_2$ concentrations as in the previous study) for calibration of uncertainties in the input data of the inversion scheme and for validation of the a posteriori emission estimates, (ii) the use of the SCIAMACHY measurements that feature much higher spatial resolution than the GOME measurements used in the previous study, (iii) independent treatment of uncertainties in anthropogenic and biogenic emissions, and (iv) extension of the inversion scheme over a much larger area including all of Europe, the Middle East, and Northern Africa. Note that the method used in our previous study was mainly designed for the relatively high level of anthropogenic air pollution typical for Western Europe, while the method proposed here is more general. Note also that the use of ozone measurements instead of $NO_2$ concentrations is advantageous for the following reasons: first, the concentration of near surface ozone is certainly more decoupled from the tropospheric $NO_2$ column than the $NO_2$ near surface concentration; this reduces the probability of error covariance in two kinds of the respective modeled data. Second, the available ozone measurements suitable for this study are much more abundant than respective measurements of nitrogen dioxide.

The paper is organised as follows. The modelled and measured data are discussed and compared in Section 2. Our inversion scheme is described in Section 3. The results of the inversion and the improvements in the model performance are discussed in Section 4. Section 5 summarises the results of this study.

## 2 Measurement and model data

### 2.1 Satellite data

We use the data for tropospheric $NO_2$ columns derived from measurements performed by the Scanning Imaging Absorption spectroMeter for Atmospheric ChartograpHY (SCIAMACHY) on board of the Envisat satellite. These data are scientific products of the University of Bremen. SCIAMACHY is a passive remote sensor covering the wavelength range from 240 and 2380 nm with a spectral resolution between 0.2 and 1.5 nm (Bovensmann et al., 1999). The $NO_2$-columns have been retrieved using measurements in the spectral window from 425 to 450 nm where the resolution of the instrument is about 0.4 nm. Although SCIAMACHY operates in both nadir and limb modes, only measurements performed in the nadir mode have been used to retrieve $NO_2$ columns for this study. The typical horizontal resolution of the instrument in the nadir mode is about $30 \times 60$ km$^2$. ENVISAT flies in a sun synchronous near polar orbit with the equator crossing time in the descending node at 10:00 a.m. local time. Global coverage is achieved after 6 days.

The method of retrieval of tropospheric $NO_2$ columns from the spectra measured by SCIAMACHY is the same as described by Richter et al. (2005). Specifically, tropospheric $NO_2$ columns are obtained in several steps, including the retrieval of slant column amounts by means of the DOAS (Differential Optical Absorption Spectroscopy) method, the estimation of the stratospheric part of the slant columns using the daily stratospheric $NO_2$ columns simulated by the 3d-CTM SLIMCAT (Chipperfield, 1999) for the time of the satellite overpass, the cloud screening (removing pixels with a cloud fraction of more than 0.2 as determined by the FRESCO algorithm (Koelemeijer et al., 2001) and the evaluation of tropospheric air mass factors (AMF) that describe the light path in the troposphere and are used to convert the tropospheric slant $NO_2$ columns to vertical columns. The vertical distribution of $NO_2$ needed for the evaluation of AMF is calculated with the MOZART-2 CTM (Horowitz et al., 2003).

Note that the use of the $NO_2$ vertical distribution simulated by MOZART (rather than CHIMERE itself) in the retrieval procedure allows us to assume that the random part of uncertainties in $NO_2$ columns derived from satellite measurements and in those modeled by CHIMERE are statistically independent. This assumption is used in our inversion procedure.

## 2.2  Data of ground-based measurements

We use measurements of near-surface ozone concentrations performed at the stations of the EMEP ground based monitoring network (http://www.nilu.no/projects/ccc/emepdata.html). Data for the period considered in this study (summer 2003) was available from 113 stations. Normally, the measurements are reported on hourly basis. However, for some stations, the data was incomplete and, therefore, some selection criteria were needed. We have taken into account only those days, for which the number of hourly measurements exceeded 16; the stations for which data were insufficient for more than 30 days have been excluded from the analysis. As a result, for the summer period of 2003 considered below, we have selected 110 stations. Unfortunately, only a few stations are located outside of Western Europe. The presence of large regions without routine ozone monitoring makes this study more difficult. Note that the important advantage of the EMEP measurements is that they are intended to reflect regional background conditions relatively unaffected by local emissions. A continental scale model can more adequately reproduce the ozone behaviour in such conditions than, for example, in big cities. The use of data from other monitoring networks (urban, regional, etc.) would seriously aggravate methodological problems concerning representativeness of measurements, data selection criteria and treatment of uncertainties.

## 2.3  Simulated columns and concentrations

The CHIMERE CTM used in this study is described in detail in the technical documentation available on the web (http://euler.lmd.polytechnique.fr/chimere/). A general description of CHIMERE can also be found in literature (see, e.g., Schmidt et al., 2001; Vautard et al., 2001, 2003; Beekmann and Derognat, 2003; Bessagnet et al., 2004). We outline below only a few major features of our modelling system.

The CHIMERE domain used in this study extends from 12°W to 65°E and from 29°N to 62°N. It covers all of Europe, the Middle East, and a part of Northern Africa with the horizontal resolution of 1°×1°. The extension of CHIMERE beyond Western Europe is discussed in Konovalov et al. (2005). In the vertical, the model has 8 layers defined with the hybrid coordinates. The top of the upper layer is fixed at the 500 hPa pressure level. Meteorological input data are calculated off-line with horizontal resolution of 100×100 $km^2$ using the MM5 non-hydrostatic meso-scale model (http://www.mmm.ucar.edu/mm5/). MM5 is initialised with NCEP Reanalysis-2 data (http://www.cpc.ncep.noaa.gov/products/wesley/ncep_data/). A simplified chemical scheme (MELCHIOR2) used in this study includes 44 species and about 120 reactions (Schmidt et al., 2001; Derognat, 2002). Lateral boundary conditions are specified using monthly average values of chemical species concentrations simulated by the MOZART model (Horowitz et al., 2003).

The data for anthropogenic emissions for $NO_x$, $SO_2$, CO, and non-methane volatile organic compounds (NMVOC) used in this study are specified using the EMEP data distributed to 11 Selected Nomenclature for Air Pollution (SNAP) sectors and gridded with the horizontal resolution of 50x50 km. The annual EMEP data for 2003 together with data provided by IER, University of Stuttgart (GENEMIS, 1994) are used to define daily, weekly, and seasonal variations of anthropogenic emissions. The data for total NMVOC emissions are split into 9 reactivity classes represented in the model mechanism taking into account typical VOC profiles for different activity sectors (GENEMIS, 1994) and following classification of hydrocarbons with regard to their structure and reactivity (Middleton et al., 1990). Note that CHIMERE does not take into account emissions from aircrafts. Such emissions are believed to provide rather insignificant fraction (1-2 %) of total anthropogenic emissions both on the global and European scales (e.g., Lee et al., 2005; Tarassón et al., 2004).

Biogenic emissions of isoprene and terpenes (the latter are affected to alpha-pinene in the chemical mechanism) are parameterised as fluxes dependent on local temperature and insolation (Guenther, 1995; Simpson et al., 1999). Parameterisation of biogenic emissions of NO from soil is based on data and methods of Simpson et al. (1999) and Stohl et al. (1996). The $NO_x$ emissions from lightning, which are probably not important for Europe, are not taken into account. The average rates of anthropogenic and biogenic emissions calculated in CHIMERE are shown in Fig. 1.

In this study, CHIMERE is used to simulate tropospheric $NO_2$ column amounts and near surface ozone concentrations for the summer months (June-August) of 2003. The choice of summer months has been motivated by the fact that this study is aimed at improving simulations of photo-oxidant pollution which is largest during the warm season. Besides, the shorter lifetime of $NO_x$ during the warm season facilitates a technical solution of the inverse modelling problem, as discussed in Section 3.4. In order to be consistent with satellite data, the modelled $NO_2$ columns are taken for each model grid cell at the time nearest to 10 A.M. of the local solar time only on those days for which satellite data are available. The daily data for the simulated $NO_2$ columns were then averaged over three summer months. Note that in contrast to tropospheric $NO_2$ columns derived from satellite measurements, $NO_2$ columns simulated by CHIMERE do not include nitrogen dioxide from the upper troposphere. Although this omission may cause systematic biases in the simulated $NO_2$ columns, it has been found earlier (Konovalov et al., 2005) that the omitted part of the tropospheric $NO_2$ can hardly contribute significantly to the spatial variability of tropospheric $NO_2$ columns. The potential systematic biases are taken into account in our inversion procedure as it is specified below (see Section 3).

When model results are used in combination with ground-based observations, it is necessary to define a model level that corresponds to a given station. The choice of the surface layer may be inappropriate for mountain sites where the model's grid cannot resolve details of a relief. We chose an appropriate model level by considering the difference between the actual height of a site (a.s.l.) and its height in the MM5 model topography.

### 2.4 Initial comparison of simulations with measurements

#### 2.4.1 $NO_2$ columns

The tropospheric $NO_2$ columns derived from the SCIAMACHY measurements in comparison with the $NO_2$ columns calculated by CHIMERE are shown in Fig. 2. As expected, CHIMERE yields somewhat smaller, on the average, $NO_2$ columns than satellite measurements. In order to eliminate systematic differences between the simulated and measured $NO_2$ columns from further analysis, we have preliminary evaluated them using a running average technique. Specifically, we first arranged the measured $NO_2$ columns for each grid cell in ascending order and then evaluated the differences of the modelled and measured $NO_2$ columns within a moving window that includes 100 grid cells. These calculations were performed separately for three big regions marked in Fig. 2b, which will be referred to by convention as Western Europe, Eastern Europe and the Afro-Asian region. The results of such analysis are shown in Fig. 3. It is seen that the differences increase with increasing magnitudes of the $NO_2$ columns. This means that the systematic errors in $NO_2$ columns are, predominantly, of multiplicative character. The considered systematic differences may include errors in both measured and modelled data. In particular, a part of the differences may be due to omission of the upper troposphere in the model, but another part may also be due to underestimations of the $NO_x$ emissions or the air mass factors used in the retrieval procedure of the satellite measurements. The modeled $NO_2$ columns with the added systematic differences are shown in Fig. 2c and the scatter plots of the measured $NO_2$ columns versus the modelled ones before and after debiasing are presented in Fig.4. Obviously, the unbiased $NO_2$ columns from CHIMERE are significantly closer to the $NO_2$ columns derived from satellite measurements than the original $NO_2$ columns calculated by the model. Nevertheless, the remaining differences, which are spatially random, are still rather considerable; a part of these differences may be due to uncertainties in emissions.

#### 2.4.2 $O_3$ concentrations

Results of the initial comparison of the simulated and measured near surface ozone mixing ratios are presented in Fig.5 for each EMEP site in terms of standard comparison statistics calculated for daily maximums of the ozone concentration. Specifically, we consider the root mean squared error (RMSE) normalised to the mean daily maximum ozone concentration at a given site and the correlation coefficient. The average values for these statistics are 0.23 (RMSE) and 0.63 (R). The correlation coefficient is higher than 0.6 for the majority of sites (73 out of 110 considered), and higher than 0.8 for 18 sites. The RMSE is smaller than 0.15 and 0.25 for 14 and 74 stations, respectively. Relatively large values of the correlation coefficient are typical for Germany, Belgium, the Netherlands, and Great Britain, but generally smaller ones are found for the sites in Eastern and Northern Europe. This observation may be indicative of the fact that CHIMERE simulates temporal ozone variability best for the sites situated in rather polluted environments where ozone behaviour is determined mainly by the photochemical processes rather than by long-range transport. Nonetheless, even when the model performs badly in terms of the correlation coefficient, it still may perform quite satisfactory with regard to normalized RMSE. For example, a very small correlation coefficient (0.17) and a rather low NRMSE (0.24) co-exist for the Shepeljovo station near Moscow. As it was argued by Konovalov et al. (2005), the quality of ozone simulations by CHIMERE is typical for state-of-the-art models. Although the calculations which we consider here were performed with lower resolution than in our previous studies (where the resolution was of $0.5° \times 0.5°$), the level of the agreement between ozone simulations and EMEP measurements practically did not change.

# 3 Inversion scheme

## 3.1 Problem formulation

As specified above, we have the datasets of the measured NO$_2$ columns ($\mathbf{C_o}$) and near-surface ozone concentrations ($\mathbf{S_o}$). We can obtain also estimates of the columns and concentrations calculated by the model as a function of emission rates, $\mathbf{C_m}$ and $\mathbf{S_m}$. The NO$_2$ columns considered here are the averages over three summer months. As to O$_3$ concentrations, we use their daily maximum values for each day during the considered summer season. The daily maximum values are expected to be more sensitive to changes of emissions than other possible characteristics of the daily ozone level (e.g., daily mean values or 8h-averages). We have also NO$_x$ emissions data (from emission inventories and the model emission interface) that are used below as a priori emission estimates ($\mathbf{E_a}$); if not specified otherwise, these are the total mean emission rates from both anthropogenic and biogenic sources over the three months. All kinds of input data may differ from their true values with unknown errors. Our main task is to elaborate a posteriori emission estimates, $\mathbf{E_p}$, that should be closer to the unknown true values of emissions than the a priori emissions. Another related task is to estimate uncertainties in the spatial distribution of a priori emissions.

## 3.2 Approach and basic assumptions

We find a solution of our inverse modelling problem using the probabilistic approach (see, e.g., Tarantola, 1987). This approach is commonly used in geophysical inverse modelling, including inverse modelling of sources of atmospheric trace gases (e.g., Enting, 2001). In accordance to this approach, we treat all considered characteristics as random values. We assume that errors in the a priori emission data are multiplicative and satisfy the lognormal probability distribution. Accordingly, the natural logarithms of true emissions, $\mathbf{e}$, satisfy the normal distribution:

$$p(\mathbf{e}|\boldsymbol{\sigma_e}) \propto \exp\left\{\sum_{i=1}^{N} -\frac{(e_i - e_{ai})^2}{2\sigma_{ei}^2}\right\}, \tag{1}$$

where $N$ is the number of grid cells in the model domain, $\mathbf{e_a}$ are the natural logarithms of the a priori emission rates, and $\boldsymbol{\sigma_e}$ is a vector of the standard deviations for errors in the a priori emissions. In the absence of more specific information, the lognormal distribution is believed to be most suitable in the case of strictly positive characteristics and multiplicative errors (Mosegaard and Tarantola, 2002). We assume also that errors of a priori emissions for different grid cells are statistically independent. Taking into account that we do not know exactly uncertainties in the a priori emissions, the standard deviations $\boldsymbol{\sigma_e}$ are also considered as random values.

In case of NO$_2$ columns, we cannot assume that errors are strictly multiplicative, because previous studies (Richter and Burrows, 2002; Martin et al., 2003; Konovalov et al., 2005) have shown that the datasets for NO$_2$ columns derived from satellite measurements also contain additive errors. Therefore, we cannot assume that the tropospheric NO$_2$ satisfy the lognormal distribution in a general case. Instead, we assume that errors in both the measured and simulated NO$_2$ columns satisfy the normal distribution with parameters dependent on the magnitude of the NO$_2$ columns.

In the context of this study, it is most important to specify the constraints that the measured NO$_2$ columns put to the emissions. Following Tarantola (1987), we combine the errors in the modeled and measured NO$_2$ columns and, as a result, we obtain the following distribution for emissions constrained by the model and measurements.

$$p(\mathbf{e}|\mathbf{C}_o,\boldsymbol{\sigma_c}) \propto \exp\left\{-\sum_{i=1}^{N}\left[\frac{(C_o^i - C_m^i(\mathbf{e}) - \delta_c^i)^2}{2\sigma_{ci}^2}\right]\right\}, \tag{2}$$

where $\boldsymbol{\delta_c}$ is the difference of the systematic errors in the measured and modeled NO$_2$ columns, and $\boldsymbol{\sigma_c}$ is the standard deviation of the combined residual errors not related to uncertainties in emissions. In the absence of certain information on the magnitudes of $\boldsymbol{\sigma_c}$, it is also considered as a random value. Note that formulation (2) assumes that the model and measurement errors are statistically independent. This would hardly be the case if our model were used in the retrieval of satellite measurements. The distributions (1) and (2) describe two different sources of information on emissions. The next reasonable step is to combine these sources in order to reduce the uncertainties in both kinds of the emission estimates. When the uncertainties in the a priori emissions and in the model and measurements are statistically independent, such combination can be performed by multiplication of the corresponding probability distribution functions. As a result, we get the following distribution for emissions that are constrained, on the one hand, by measurements, and, on the other hand, by data of emission inventories:

$$p(\mathbf{e}|\mathbf{C}_o,\boldsymbol{\sigma_c},\boldsymbol{\sigma_e}) \propto \exp\left\{-\sum_{i=1}^{N}\left[\frac{(C_o^i - C_m^i(\mathbf{e}) - \delta_c^i)^2}{2\sigma_{ci}^2} + \frac{(e^i - e_a^i)^2}{2\sigma_{ei}^2}\right]\right\} \tag{3}$$

Note that the same distribution (3) could also be obtained using Bayes's theorem.

If we knew the probability distribution functions (PDF) for $\sigma_c$ and $\sigma_e$, we could integrate (3) with the weights defined by those distributions. In doing so, we would eliminate the unknowns from the PDF for emissions. However, as we lack information on the distributions of $\sigma_c$ and $\sigma_e$, we use another approach. Specifically, we consider a posteriori emission estimates that provide a maximum of the distribution (3) under the given values of the standard deviations. These maximum likelihood estimates, $\hat{\mathbf{e}}$, satisfy the following equations:

$$\sum_{i=1}^{N}\left[(C_o^i - C_m^i(\hat{\mathbf{e}}) - \delta_c^i)\frac{\partial C_m^i(\hat{\mathbf{e}})}{\sigma_{ci}^2 \partial \hat{e}^j}\right] + \frac{1}{\sigma_{ej}^2}(\hat{e}^j - e_a^j) = 0, \qquad j=1,\ldots,N. \tag{4}$$

The standard deviations are estimated based on comparison of model results with observational data as described in the next section.

### 3.3 Uncertainties in input data

Let us consider the following variance which is defined for a given grid cell "$i$" as follows:

$$\Delta_{Ti}^2 = \frac{1}{N_w}\sum_{j=1}^{N_w}\left(C_o^{k(i,j)} - C_m^{k(i,j)}(\mathbf{e}_a) - \overline{C}_o^i + \overline{C}_m^i(\mathbf{e}_a)\right)^2 \tag{5}$$

where $N_w$ is the number of data points in the "window" comprising preselected grid cells with the magnitudes of the measured $NO_2$ columns similar to that in the grid cell "$i$", $\overline{C}_o^i$ and $\overline{C}_m^i$ are the mean values of the observed and modeled $NO_2$ columns in the given window, $k$ is the index of grid cells inside the model domain and $j$ is the internal index of a grid cell inside the window. Obviously, when $N_w$ is sufficiently large, this variance characterizes total random (in the spatial sense) unbiased errors in the modeled and measured $NO_2$ columns. Similarly, the standard deviation $\sigma_c$ that characterizes a combination of the measurement error and a part of the model error, which is not caused by random uncertainties in emissions, relates to differences between the modeled and measured columns as follows:

$$\sigma_{ci}^2 \cong \frac{1}{N_w}\sum_{j=1}^{N_w}\left(C_o^{k(i,j)} - C_m^{k(i,j)}(\mathbf{e}_c) - \overline{C}_o^i + \overline{C}_m^i(\mathbf{e}_c)\right)^2, \tag{6}$$

where $\mathbf{e}_c$ are the logarithm of emissions that could be obtained after correcting all random errors in a priori emissions. Finally, the model errors due to uncertainties in the spatial distribution of $NO_x$ emissions are characterized by the following variance:

$$\Delta_{ei}^2 = \frac{1}{N_w}\sum_{j=1}^{N_w}\left(C_m^{k(i,j)}(\mathbf{e}_a) - C_m^{k(i,j)}(\mathbf{e}_c)\right)^2. \tag{7}$$

Assuming that the parts of the error in the modeled $NO_2$ columns that relate and do not relate to uncertainties in emissions are statistically independent, we get the following approximate relation:

$$\sigma_{ci}^2 \cong \Delta_{Ti}^2 - \Delta_{ei}^2 \tag{8}$$

Given $\Delta_e$ and $\Delta_T$, the relation (8) provides the estimate of the parameter $\sigma_c$. However, while the variance $\Delta_T$ is evaluated directly using measured and modeled data, the variance $\Delta_e$, which quantifies the differences between $NO_2$ columns calculated with a priori and corrected (unknown) emissions, is to be estimated approximately. In order to better understand the meaning of $\Delta_e$, let us first assume that uncertainties in emissions in different grid cells are statistically independent and that the nonlinearity of the relationship between $NO_2$ columns and $NO_x$ emissions may be neglected. After linearizing Eq. (7), we obtain the following relation between $\Delta_e$ and $\sigma_e$:

$$\Delta_{ei}^2 \cong \frac{1}{N_w}\sum_{j=1}^{N_w}\sum_{n=1}^{N}\left(\frac{\partial C_m^{k(i,j)}}{\partial \hat{e}^n}\right)^2 \sigma_{en}^2. \tag{9}$$

This result demonstrates that when $\Delta_e$ is estimated for a sufficiently large statistical ensemble, its magnitude depends mainly on the mean deviations of the a priori emissions from the true ones, rather than on the true emission explicitly. Accordingly, an idea of our estimation of $\Delta_e$ is to substitute the unknown values of $\mathbf{e}_c$ in Eq. (7) for some surrogate values defined such that the differences between the natural logarithms of the surrogate and a priori emissions would have the variance equal to $\sigma_e^2$ (that is, the same as the variance of the unbiased differences between the natural logarithms of the true and a priori emissions). Taking into account that emission uncertainties in neighboring grid cells may be not completely independent, it is also important that the spatial structure (covariances) of the differences between the logarithms of the surrogate and a priori emissions would be similar to the spatial structure of the uncertainties in a priori emissions. In order to satisfy these conditions and taking into account that the best available estimates of the true emissions are the a posteriori estimates $\hat{\mathbf{e}}$, we define the surrogate emissions, $\mathbf{e}_s$, by scaling a posteriori emissions as follows:

$$e_s^i = \frac{\left\{\sum_{j=1}^{N_w}\left(\sigma_e^{k(i,j)}\right)^2\right\}^{1/2}(\hat{e}^i - e_a^i)}{\left[\sum_{j=1}^{N_w}\left(\hat{e}^{k(i,j)} - e_a^{k(i,j)}\right)^2\right]^{1/2}} + e_a^i. \quad (10)$$

Note that the a posteriori emissions $\hat{\mathbf{e}}$ could not be used directly as a substitute for $\mathbf{e}_c$ because the root mean squared differences between $\hat{\mathbf{e}}$ and $\mathbf{e}_a$ are smaller than the presumed uncertainties in the a priori emissions. On the other hand, we could not replace $\mathbf{e}_c$ by randomly perturbed a priori emissions because in that way the probable spatial covariances of errors in the a priori emissions would be ignored. This could eventually lead to overestimation of $\sigma_c$ and $\sigma_e$.

After substituting $\mathbf{e}_c$ in (7) for $\mathbf{e}_s$, we get the following approximate relationship between $\sigma_c$ and $\sigma_e$:

$$\sigma_{ci}^2 \cong \Delta_{Ti} - \frac{1}{N_w}\sum_{j=1}^{N_w}\left(C_m^{k(i,j)}(\mathbf{e}_a) - C_m^{k(i,j)}(\mathbf{e}_s[\sigma_e,\sigma_c])\right)^2 \quad i=1,\ldots,N \quad (11)$$

Note that since $\mathbf{e}_s$ depends on the solution $\hat{\mathbf{e}}$ of Eq. (4), they depend both on $\sigma_c$ and $\sigma_e$.

We then take into account that the uncertainty in a priori emissions may depend on the magnitude of the emission rate in a given grid cell. In order to specify this dependence, we consider the differences between two different emission inventories, such as EMEP and POET (http://www.aero.jussieu.fr/projet/ACCENT/POET.php) for anthropogenic emissions, and also between soil $NO_x$ emissions calculated in CHIMERE and provided by the GEIA emission inventory (Yienger and Levy II, 1995). Both the EMEP and POET emissions were taken for the year 2000, since the POET $NO_x$ emissions were not available for more recent years. We define the logarithmic deviation $\mathbf{D}$ between the data $\mathbf{e}_1$ and $\mathbf{e}_2$ of a pair of different emission inventories as follows:

$$D_i^2 = \frac{1}{N_w}\sum_{j=1}^{N_w}\left(e_1^{k(i,j)} - e_2^{k(i,j)}\right)^2 \quad (12)$$

Here we use again a windowing technique which involves the preselection of $N_w$ grid cells with similar magnitudes of emission rates. Values of $D$ calculated for both anthropogenic and biogenic emissions and for different regions are shown in Fig. 6. Assuming that the covariance of errors in data of different emission inventories are positive, these deviations can be considered as the lower limit of total uncertainties in both kinds of data. Evidently, that these uncertainties are quite considerable. It is interesting to note that somewhat in contrast to our intuitive expectations, the considered differences between emission data are not always smaller for Western Europe than for the other two regions. However, while interpreting the dependencies shown in Fig.6, it is necessary to take into account that values of $D$ may be larger for one region than for another simply due to larger covariance of uncertainties in data of different emission inventories for the first region. The decreasing character of the dependence of $D$ on the magnitude of the emission rate indicates that the relative uncertainty in emissions is probably largest for grid cells with small emission rates, while it is smallest for grid cells with large emission rates. We combine the uncertainties in anthropogenic and biogenic emissions as follows:

$$\sigma_{ei}^2 = \frac{D_{ai}^2 E_{ai}^2 \xi_a^2 + D_{bi}^2 E_{bi}^2 \xi_b^2}{(E_{ai} + E_{bi})^2}, \quad (13)$$

where $D_a$ and $D_b$ are the logarithmic deviations defined by Eq. (15) for anthropogenic and biogenic emissions, respectively, and

$$\xi_{a,b}^2 = \frac{\sigma_{a,b}^2}{N^{-1}\sum_{i=1}^{N}D_{ai}^2} \quad (14)$$

are scaling factors defined with given average (for the whole region considered) values $\sigma_a$ and $\sigma_b$ of the standard deviations of a priori anthropogenic and biogenic emissions, respectively. $N$ is the total number of grid cell considered for a given region. After substituting Eqs. (13) and (14) into Eq. (11), the latter enables estimation of $\sigma_c$ as a function of two yet unknown scalar parameters $\sigma_a$ and $\sigma_b$.

Finally, we try to constrain values of $\sigma_a$ and $\sigma_b$ based on the idea that their optimal values should yield the best agreement between the model with a posteriori emissions and some independent measurements that have not been used for fitting the a posteriori emissions. In this study, we use measurements of near-surface ozone. It seems obvious that the uncertainties in the measured data for ozone concentrations and tropospheric $NO_2$ columns can be considered as independent. Besides, the covariance of uncertainties in the modeled ozone concentrations and $NO_2$ columns is likely also small, because the behavior of these characteristics is driven by rather different physico-chemical processes. Since both the measured and modeled ozone concentrations are, in our case, strictly positive, it is reasonable to assume that distributions of measurement and model errors in ozone concentration satisfy the lognormal probability distribution. Then, using Bayes's theorem, we can get the following probability distribution for $\sigma_a$ and $\sigma_b$:

$$p(\sigma_a, \sigma_b | \mathbf{s}_o) \propto \exp\left\{-\sum_{k=1}^{L}\sum_{m=1}^{N_k}\left[\frac{(s_o^{mk} - s_m^{mk}(\bar{\mathbf{e}}[\sigma_a, \sigma_b]) - \overline{s}_o^{mk} + \overline{s}_m^{mk})^2}{2\sigma_{sk}^2}\right]\right\} \quad (15)$$

where $\mathbf{s}_o$ and $\mathbf{s}_m$ are the logarithms of the observed and modeled daily maximums of ozone concentrations, $k$ is the index of a monitoring station, $N_k$ is the number of daily data for a station "$k$", and $\sigma_s$ is the standard deviation of the unbiased combined error in the measured and modeled ozone concentrations. Here we assumed also that the a priori distributions of the logarithms of $\sigma_a$ and $\sigma_b$ are uniform (that is, there are no a priori constraints to their values). The maximum-likelihood (optimal) estimates of $\sigma_a$ and $\sigma_b$ should yield a minimum of the corresponding cost function, i.e. we are looking for a combination of $\sigma_a$ and $\sigma_b$ that yields a posteriori emissions in a way to minimize differences in the modeled and observed surface ozone:

$$G = \sum_{k=1}^{L}\sum_{i=1}^{N_k}\left[\frac{(s_o^{ik} - s_m^{ik}(\bar{\mathbf{e}}[\sigma_a, \sigma_b]) - \overline{s}_o^{ik} + \overline{s}_m^{ik})^2}{\sigma_{sk}^2}\right]. \quad (16)$$

It seems reasonable to assume that the combined error in ozone concentration is mostly not related to uncertainties in emissions. This assumption, which is found to be consistent with the results of this study, allows us to estimate $\sigma_s$ as follows:

$$\sigma_{sk}^2 \cong \frac{1}{N_k}\sum_{i=1}^{N_k}(s_o^{ik} - s_m^{ik}(\mathbf{e}_a) - \overline{s}_o^{ik} + \overline{s}_m^{ik})^2 \quad (17)$$

Taking into account Eqs. (11) and (13), the minimization of the function $G$ (see Eq. (16)) yields optimal values of $\sigma_a$ and $\sigma_b$ along with the consistent estimates of uncertainties in $NO_2$ columns $\sigma_c$. In other words, while equation (11) gives an estimation of $\sigma_c$ under the given values of $\sigma_a$ and $\sigma_b$ (after substituting the relation (13) which translates overall a priori emission uncertainties into emission uncertainties for individual grid cells), the optimization of the agreement between modeled and measured ozone concentrations (Eq. (16)) gives the best estimates of overall uncertainties in a priori anthropogenic and biogenic emissions.

Unfortunately, due to the lack of ground-based observations outside of Western Europe, the method described above cannot be fully applied to Eastern Europe and the Afro-Asian region. On the other hand, we cannot expect that the uncertainties in emissions are the same for all regions. Therefore, some less obvious assumptions about emission uncertainties outside of Western Europe are unavoidable. Specifically, we assume that the scaling factors $\xi$ (see Eq. (14)) are the same for all regions. This assumption implies, for example, that if the EMEP data for Western Europe were found less uncertain than the POET data, they would also be expected to be less uncertain in the other regions. If the covariance of errors in the considered emission datasets is larger for the regions outside of Western Europe (that seems to be probable in view of results shown in Fig. 6), such an assumption may lead to underestimation of emission uncertainties in Eastern Europe and the Afro-Asian region. Therefore, we consider our estimates of emissions uncertainties in these regions as the lower limit of probable real uncertainties. Finally, we assume that the uncertainties in $NO_2$ columns $\sigma_c$ in Eastern Europe and in the Afro-Asian region are the same as in Western Europe.

### 3.4 Numerical method

First of all, it is necessary to note that finding a mathematically precise solution of the inversion problem coupled with the problem of the estimation of uncertainties would, in our case, be extremely computationally expensive, taking into account that a single three-month run of our model takes about 10 hours on a standard PC. Therefore, we use approximate methods. The core of our method is the calculation of the modeled relationships between $NO_x$ emissions and $NO_2$ columns or concentrations by substituting the original model with a set of linear statistical models describing the considered relationships approximately. In doing so, we assume that the typical range of transport of freshly emitted $NO_x$ is rather limited. This approximation was tested and used in our earlier study (Konovalov et al., 2006). Specifically, our statistical models are defined as follows:

$$C_i(\mathbf{E}) := C_i(\mathbf{E}_a)\left(1 + \sum_{j=1}^{(2M+1)^2} \alpha_j^i \left[\exp(e_j^i - e_{aj}^i) - 1\right]\right), \quad (18)$$

where $\alpha_j^i$ are regression coefficients, which represent the sensitivity of the $NO_2$ column in the grid cell $i$ to a perturbation of the emission rate in the grid cell $j$, and $M$ is the number of layers of the grid cells around the "central" cell $i$: the larger $M$ is, the more distant transport of $NO_x$ is taken into account. The models are built using results of a sufficiently large number of model runs with pre-specified small random perturbations of emissions in each of the model grid cells. Accordingly, the coefficients $\alpha_j^i$ are obtained by solving a set of linear equations for relative perturbations of emission rates and corresponding responses of $NO_2$ columns. Technically, the respective equations are solved using the SVD method (Press et al., 1992). The optimal number of "random" model runs depends on the desired accuracy, the effective distance of transport, and the nonlinearities in the real relationships between emissions and columns, but, importantly, it does not depend on the total number of grid cells as soon as the dimensions of the model domain exceed the typical range of $NO_2$ transport. Similar to our previous study, we found it to be sufficient to consider the models with M equal 2 that have been built based on results of 100 model runs. Note that although this number is not small, it is, nevertheless, much smaller than the number of the emission parameters to be optimised, and, therefore, our method is very advantageous in terms of the computational demand when compared with the direct variational approach. Special tests described in Konovalov et al. (2006) show that, if the input data were perfect, our method could enable significant (up to 3 times) reduction of uncertainties in a priori emissions, while the maximum reduction of uncertainty with real (noisy) data was limited by uncertainties in input data.

The numerical scheme includes several embedded optimisation procedures, which are outlined in Fig. 7. The optimisation procedure of the lowest level is aimed at finding a solution of Eq. (4) under given values of the standard deviations $\sigma_c$ and $\sigma_e$. The solution is found by means of the steepest descent method using the a priori emissions as initial guess. Although the steepest descent method is not optimal in terms of the computational demand, it has been chosen because of its robustness and stability. One hundred iterations that are found to be sufficient to reach the minimum require only several seconds of CPU time. The optimisation procedure of the next level is intended for estimation of $\sigma_c$ (by solving Eq. (11)) under given values of $\sigma_a$ and $\sigma_b$. The solution is found by means of the simple iteration method with "damping" when values of $\sigma_c$ obtained in the left-hand side of Eq. (11) are used at the next iteration to construct values $\sigma_c$ in the right hand side of the equation. This procedure requires also about one hundred iterations. The optimization procedure of the highest level is intended for estimation of $\sigma_a$ and $\sigma_b$ using ozone observations by means of minimization of the cost function $G$ (see Eq. (16)). Since the relations between $NO_x$ emissions and near surface ozone concentrations are essentially nonlinear and non-local, the repeated runs of the whole model are inevitable at this step. Considering the huge computational coasts of the exact solution of this two-dimensional minimization problem (note that we need a global minimum, while most of the "automatic" methods only ensure finding a local minimum closest to an initial guess), we find an approximate solution using a kind of a simplest "global search" method. Specifically, we perform an ensemble of model runs with a posteriori emissions corresponding to sets of fixed values of $\sigma_a$ and $\sigma_b$. The set of values of $\sigma_a$ includes the numbers from zero to 0.9 with equidistant steps of 0.1, while the set of values of $\sigma_b$ consists of the following numbers: 0, 0.1, 0.3, 0.6, 0.9, 1.2, 1.6, 2.0, 2.5, 3.0. We considered only such combination of all these possible values of $\sigma_a$ and $\sigma_b$ that satisfy to the condition $\sigma_a < \sigma_b$, as it seems reasonable to expect that uncertainties of biogenic emissions are larger than uncertainties in anthropogenic emissions.

## 3.5 The Monte Carlo experiment

An inverse modeling study is traditionally expected to provide not only optimized a posteriori emissions but also estimates of their uncertainty. Besides, as soon as our estimates of uncertainty in a priori emissions may also be regarded as an important result of our study, it is useful to evaluate the accuracy of these estimates. In order to do all this, we employ the Monte Carlo method (see, e.g., Press et al. 1992). Specifically, we use the a posteriori emissions and corresponding modelled $NO_2$ columns as surrogates for the true data and generate a number of sets of synthetic data

$$e_{as} = e_p + \Delta_e,$$
$$c_{os} = c_m(e_p) + \Delta_c, \qquad (19)$$

for the a priori emissions and the measured $NO_2$ columns with "errors" $\Delta_e$ and $\Delta_c$ sampled from the probability distributions (1) and (2). After performing inversion with so defined synthetic data and considering the differences with the optimal a posteriori emissions, it would be easy to evaluate uncertainties in a posteriori emissions due to the presumed uncertainties in the $NO_2$ columns and the a priori emissions. However, we should take into account that our knowledge of uncertainties in the input data is also not quite perfect. For example, an underestimation of the uncertainty in the a priori emissions would result in smaller corrections of the a priori emissions than it is actually required. And vice versa, the overestimation of the uncertainty in the a priori emissions would result in unrealistically large corrections to a priori emissions and attribution of part of the model and measurement error to uncertainties in emissions. Thus, in our opinion, this source of potential uncertainties in the inverse modelling results should also be taken into account.

Because we estimate the uncertainty parameters of distributions (1) and (2) by optimising the agreement between the measured and modelled near surface ozone concentration, the uncertainty in these estimates may relate to the facts that (i) both the measured and modelled ozone data may be rather inaccurate, (ii) the amount of available ozone measurement data is always limited, and (iii) the spatial distribution of the a posteriori emissions is not perfect. Ideally, these sources of uncertainties could be taken into account in the same way as the uncertainties in $NO_2$ columns and $NO_x$ emissions. However, this way would be computationally too expensive, as it would require repeating the whole inversion procedure many times. Instead, we employ a simpler and more robust approach, which is based on the boot-strapping method (see, e.g., Efron et al., 1993). Namely, we repeat all the procedures described above but with randomly selected subsets of ozone monitors. Each subset included one third of the total number of ozone monitors in Western Europe. Therefore, in each experiment we deal with three independent subsets of ozone data. While systematic uncertainties in ozone measurement or modelled data are removed from analysis by debiasing (see Eq. (16)), it seems reasonable to expect that random uncertainties associated with each subset of ozone data are independent. Besides, as the location of different stations is different, the uncertainties in the spatial distribution of the a posteriori emissions can also be efficiently taken into account. The corresponding estimates of the standard deviations $\sigma_a$ and $\sigma_b$ of the a priori emissions are used (i) to assess the uncertainty in optimal estimates of $\sigma_a$ and $\sigma_b$ obtained with the whole dataset of ozone measurements and (ii) to generate "synthetic" data (19). These synthetic data corresponding to different subsets of ozone monitors are then combined and analysed in order to get final estimates of uncertainties in the a posteriori emissions. It should be noted that because each independent estimate of $\sigma_a$ and $\sigma_b$ in the Monte Carlo experiment is based on a considerably smaller amount of ozone data than the main estimation, it is very probable that the approach used here overestimates the uncertainties $\sigma_a$ and $\sigma_b$. In turn, this overestimation may result in overestimation of uncertainties in a posteriori emissions. Therefore, we consider our estimates of uncertainties in a posteriori emissions and related parameters as the upper limits of their actual uncertainties.

## 4 Results

### 4.1 Estimates of uncertainties in input data

Let us first consider the estimates of uncertainties in a priori emissions obtained as a result of the optimisation procedure described above. The average values and uncertainties of the logarithmic standard deviations of the a priori emissions for different regions are given in Table 1. These estimates correspond to the scaling factors $\xi_a$=0.42 and $\xi_b$=0.88 (see Eq. (14)). That is, the estimated uncertainties in both anthropogenic and biogenic emissions used in CHIMERE are, on average, smaller than the differences between data of different emission inventories. Note that the reported uncertainties in $\sigma_a$ and $\sigma_b$ for Eastern Europe and the Afro-Asian region should be considered as the lower limit of the actual uncertainties, because our estimates are based of the assumption that the scaling factors $\xi_a$ and $\xi_b$ for these regions are the same as for Western Europe. This assumption results, in particular, in significantly lower values of $\sigma_b$ for Eastern Europe and the Afro-Asian region than for Western Europe. It is useful to take also into account that $\sigma_a$ and $\sigma_b$ are intended to characterize only the random part (in the spatial sense) of the total uncertainties in the a priori emissions. Because the structure of the land use is likely to be more homogeneous in less urbanized regions outside of Western Europe, the fraction of the systematic errors in emissions may be larger there at the expense the random errors. As to the presumable mean uncertainties in the anthropogenic emissions ($\sigma_a$), they are practically the same for all regions. As it is noted above, it is possible that values of $\sigma_a$ for Eastern Europe and the Afro-Asian region used in this study are underestimated. The underestimation of $\sigma_a$ (as well as of $\sigma_b$) may result in smaller corrections for a priori emissions in the respective region than actually needed.

The estimated values of the standard deviation $\sigma_c$ for uncertainties in $NO_2$ columns are shown in Fig. 8. For comparison, we show values of the variance $\Delta_T$ that characterizes the total random uncertainty in $NO_2$ columns (including the uncertainties in the modeled $NO_2$ columns due to errors in a priori emissions). We also show values of the variance $\Delta_e$, which estimates the part of the uncertainties in the model due to uncertainties in a priori emissions. As it is follows from our results, the magnitudes of uncertainties in $NO_2$ columns caused by errors in emissions are very similar to the magnitudes of other uncertainties. Regarding the dependence of $\sigma_{ci}$ on the magnitude of the $NO_2$ columns, it is useful to note that this dependence reveals the presence of the additive and multiplicative parts of uncertainties in $NO_2$ columns. The additive error is evidently about $2 \cdot 10^{14}$ cm$^2$, while the multiplicative error is between 10 and 20 percent.

### 4.2 A posteriori emissions

Figure 9 presents the ratios of the a posteriori to a priori emissions and also the differences between a posteriori and a priori emissions. These correction factors and differences were obtained with optimal estimates of $\sigma_a$, $\sigma_b$ and $\sigma_c$ discussed above. Considering the results shown in Fig. 9, it is evident that the obtained emission corrections are not distributed truly randomly in space. On the contrary, the positive and negative changes of the a priori emissions in neighbouring grid cells tend to covariate. In particular, in accordance to our results, the a priori emissions are probably overestimated for the largest part of Great Britain, Belgium, the Netherlands, Lithuania, Latvia, Poland, Bulgaria, Greece, and Iraq, but mostly underestimated for Spain, Italy, Switzerland, Czechia, Former Yugoslavia,

Turkey, Lebanon and Iran. The emission correction factors are typically inside of the range from 0.5 to 2, but there are also exceptions. Regarding the possible origin of these covariances, it is useful to note first that we revealed similar covariances (although not exactly for the same regions) when we considered spatial distributions (not shown here) of ratios (or differences) of the different emission data used in this study for approximating the dependence of emission uncertainties on the magnitude of the emission rate in a given grid cell (see Section 3.3). This observation means that the regional covariances in emission uncertainties can actually exist. Most probably, they are a consequence of systematic biases in emission factors that were used for estimation of $NO_x$ emissions from predominant (for a given region or country) types of sources. However, it should be taken into account that the model/measurement errors may also spatially covariate due to specific features (meteorology, land use, topography) of different regions. And besides, the inverse modelling scheme based on the probabilistic approach (as well as any inversion scheme involving regularization) may itself introduce some smoothing to the real distribution of errors in emissions. Because of these possibilities, one cannot be sure that the a posteriori emission estimates are actually more accurate than a priori emissions for individual pixels. Nonetheless, as it is argued below, on the average, the uncertainty in a posteriori emissions is smaller than the uncertainty of the a priori emissions. In general, the emission correction factors provided here should be considered in the probabilistic sense together with estimates of their uncertainties.

The estimates of uncertainties in the a priori and a posteriori total emissions and the relative reduction of the uncertainties are shown in Fig. 10. This figure provides two levels of information. The first level is the uncertainty values depicted in colour, while the second level is the fact that the pixel is not blank. Because the uncertainty in anthropogenic emissions is found to be significantly lower than the uncertainty in biogenic emissions (see Table 1), the uncertainty in total emissions is typically lower in the regions where the fraction of anthropogenic emissions is larger. It is useful to note also that in accordance to the assumption about statistical independence of uncertainties in anthropogenic and biogenic emissions, the uncertainty in the total emission rate in a given grid cell may be smaller than the uncertainties in either anthropogenic or biogenic emission rates taken separately. In several grid cells, the confidence interval for a posteriori estimates turned out to be larger than the confidence interval for a priori emissions. This is mainly because of possible uncertainties in the internal parameters of the inverse modelling scheme (such as $\sigma_c$ and $\sigma_e$), Nonetheless, the results for such grid cells are not necessarily useless, as long as the sign of the difference between a posteriori and a priori emissions in a considered grid cell is statistically significant. In order to distinguish between these possibilities, the relative reduction of the uncertainty in a priori emissions is presented in Fig. 10c only for those grid cells for which it actually takes place (i.e., it is positive), while Figure 10b shows additionally results for those grid cells for which the direction of the proposed change in a priori emissions is statistically significant (in terms of the 68.3 percentile). The statistical significance of the direction of the a priori emission change is also estimated in the Monte Carlo experiment. The averages of a priori and a posteriori uncertainties and of the uncertainty reductions for different regions are listed in Table 2. Although the uncertainty reduction is, on average, not large, it is significant for many grid cells, especially in Western Europe where the reduction exceeds 30 percent almost for each fourth grid cell among those for which the emission uncertainty is reduced. The reduction of uncertainties is, on average, smaller outside of Western Europe, but still is considerable (more than 20 percent) for most of the major cities (with population of about 1 million or more). We would like to remind that the a posteriori uncertainties reported here are likely to be overestimated (see Sec. 3.5). Accordingly, it is probable that the uncertainty is actually reduced for a larger number of grid cells than shown in Fig. 10c. Because of this possibility, we do not exclude any grid cells from the a posteriori emission dataset when using it for checking the agreement of the modelled results with measurements (see section 4.3). Otherwise, it is always possible to replace the a posteriori emissions in the grid cells where their uncertainty is expected to be too large with the a priori emissions.

Table 3 lists the a priori and a posteriori emission rates for 54 major cities marked on our plots. Most of these estimates are obtained using bilinear interpolation between four grid cells closest to the city centre. Naturally, such estimates reflect not only emissions within the city boundaries but in most cases, also emissions from "satellite" cities. Nevertheless, we believe that such data are useful, particularly because they are rather easy to use for inter-comparison with similar results of other inverse modelling studies and emission inventories. In accordance to our results, the a priori emissions are probably overestimated (or underestimated) for half of the cities (27) listed in Table 3. However, the differences between the a priori and a posteriori emissions exceed the uncertainty range only for part of the cities (21), which are marked in bold in Table 3. In particular, a statistically significant overestimation of a priori emission is found for Baghdad (a factor of 3.1), Bucharest (1.3), Copenhagen (1.4), Enteb (1.3), London (1.2), Nizhny Novgorod (1.3), S.- Petersburg (1.8), Samara (1.3), Stockholm (1.3) and Volgograd (1.5), while an underestimation is probable for Amman (1.5), Beirut (3.8), Bursa (1.7), Isfahan (2.1), Istanbul (1.7), Izmir (1.7), Madrid (1.3), Tabriz (6.3), Tbilisi (3.2), Tehran (3.3) and Yerevan (3.4).

It is useful to note that the spatial distribution of uncertainties in a priori emissions for Western Europe is, overall, in reasonable agreement with the results of our previous study (Konovalov et al., 2006). However, the different spatial resolution of the model domains used in these two studies and different years of the analysis prevent us from direct comparison of the results.

### 4.3 Validation of a posteriori emissions with measurement data

It seems reasonable to expect that if the a posteriori emissions are really better, they should lead to improvements in the agreement between the modelled and measured data. The $NO_2$ columns calculated with the a posteriori emissions are compared with the $NO_2$ columns derived from satellite measurements in scatter plots shown in Fig. 11. Similar scatter plots but obtained with the a priori emissions were discussed above (see Fig. 4). The comparison of results presented in Figs. 4 and 11 reveals that the $NO_2$ columns calculated with the a posteriori emissions demonstrate much better agreement with the measured $NO_2$ columns. In particular, the coefficient of determination ($R^2$) calculated for the $NO_2$ columns from SCIAMACHY and the unbiased $NO_2$ columns from CHIMERE is increased from 0.88 to 0.94. This means, that the residual spatial variability, which is not taken into account by the model, is reduced by a factor of 2. The agreement still is not perfect, particularly because a considerable part of the uncertainties in $NO_2$ columns is not due to uncertainties in emissions (see Fig. 8). Anyway, this result proves that the inversion scheme works properly. However, the improvement of the agreement between the measured and modelled $NO_2$ columns can only be considered as necessary but still insufficient evidence in favour of our emission estimates. Indeed, a very good agreement between measured and modelled $NO_2$ columns could, in principle, be insured by assigning unrealistically large values for the uncertainties in a priori emissions and/or too small values for the uncertainties in $NO_2$ columns. But in such a case, the differences between the a posteriori and a priori emissions would mostly reflect uncertainties in the available data for $NO_2$ columns rather than actual uncertainties in the a priori $NO_x$ emissions.

The comparison of the modelled and measured near surface concentrations of ozone provides a more critical test for our a posteriori emissions, since these data have not been used directly for fitting $NO_x$ emissions, but only for optimization of two "global" parameters ($\sigma_a$ and $\sigma_b$) of the inversion scheme. While the results discussed here are obtained with optimal values of these parameters, it is important to note that the comparison statistic discussed below are not very sensitive to their changes. In fact, we have found that if the use of a posteriori emissions improves some ozone statistics, the improvement takes place in a very wide range of values of $\sigma_a$ and $\sigma_b$. Moreover, values of $\sigma_a$ and $\sigma_b$ that correspond to the minimum of the cost function G (see Eq. (16)) do not necessarily yield optimum values of other possible ozone comparison statistics.

Figure 12 presents the scatter plots of the seasonally average daily maximums of ozone concentrations at the stations in Western and Eastern Europe. While considering these results, it is useful to keep in mind that the spatial distribution of $NO_x$ emissions is only one of the numerous factors controlling the level of near surface ozone. Among other factors, there are, in particular, the deposition, the vertical and horizontal transport, chemical transformations of a number of related species, spatial distribution and temporal evolution of emissions rates of the spectrum of hydrocarbons. As neither of these processes is simulated perfectly, it is reasonable to expect that even if we would be able to correct all errors in the spatial distribution of the seasonally mean $NO_x$ emissions used as the a priori emissions in our model, this would not result in a strong reduction of uncertainties in the simulated ozone concentrations. Moreover, measurement sites may not be entirely representative for a model grid cell. Nevertheless, the results presented in Fig. 12 demonstrate the evident improvement of the agreement between ozone simulations and measurements. In particular, the coefficient of the determination (square of the correlation coefficient) is increased by 21 percent (from 0.37 to 0.45) for Western Europe and 7 (from 0.53 to 0.57) percent for Eastern Europe. In order to check that the attained level of the improvement of the agreement between the simulated and measured ozone concentrations is not inconsistent with the estimated level of the uncertainty in the a priori emissions, we considered a test case for which we used the perturbed a priori emissions with additional uncertainties sampled randomly from the lognormal distributions. The level of these additional uncertainties corresponded to our estimates of uncertainties in the a priori emissions. Using the perturbed emissions, we found the coefficient of the determination to decrease only by 10 percent for Western Europe and 11 percent for Eastern Europe. This decrease gives the order of improvement that we can expect when replacing a priori emissions by perfect emissions. Note that in our test we disregarded possible spatial covariances of the emission uncertainties; this can explain why the decrease in the coefficient of determination for Western Europe in the test case is smaller than its increase with the a posteriori emissions.

It seems reasonable to expect that the sensitivity of ozone to changes of emissions is larger for days on which the photochemical ozone production is strong. Such days should correspond to high percentiles of ozone concentrations. Figure 13 shows the scatter plots of $90^{th}$ percentile of observed and modelled daily maximums of ozone concentration. Because $90^{th}$ percentiles are at the edge of statistical distribution, the overall agreement between the model and observation is as expected worse here than in the case with the seasonally average ozone concentrations. However, the relative improvements due to corrections in $NO_x$ emissions are larger. In particular, the coefficient of determination is increased from 0.13 to 0.22 (that is, 61 percent) for Western Europe and from 0.44 to 0.54 (22 percent) for Eastern Europe.

We also considered the relative differences in the root mean squared errors (RMSE) and the differences in the linear correlation coefficient calculated for each station with the a priori and a posteriori emissions. Both statistics were calculated for daily maximums of ozone concentrations. It is found that improvements take place for most of the stations. Specifically, the root mean squared error is reduced for 74 out of 110 stations and the correlation coefficient is increased for 69 stations. A simple statistical test (based on a Monte Carlo experiment) shows that if the result of the inversion were equivalent to a random change of the comparison statistics for the considered set of the stations, then the probability of having the correlation coefficient increased at 69 or more stations would be only about 1 percent. Considering the ozone statistics separately for three different regions defined above we find that both RMSE and the correlation coefficient is improved for 62 stations out of 94 at Western Europe. In Eastern Europe, RMSE is improved for 11 stations out of 15, but the correlation coefficient is improved only for 6 stations. Both statistics are improved for a single station at the Afro-Asian region. Because the number of stations located outside of Western Europe is rather limited, the results obtained for Eastern Europe and the Afro-Asian regions alone are not statistically significant. In particular, although the changes of the correlation coefficient for the stations at Eastern Europe seem to indicate that the a posteriori emissions are worse than the a priori emissions, the probability that the hypothesis about entirely random origin of this result is false is less than 40 percent. Therefore, considering all results, we can conclude that the use of the a posteriori emissions in CHIMERE results in statistically significant improvements of ozone simulations both for the whole ensemble of EMEP stations and for the stations located only in Western Europe. Taking into account that the level of near surface ozone concentrations and tropospheric column amounts of nitrogen dioxide are driven by significantly different processes, we consider this result as a good argument that the emission estimates elaborated in this study are actually more accurate than the emissions used in the standard version of our model. In future, it would be useful also to try to employ the a posteriori emissions elaborated in this study with a different model. The respective emission data (in the digital form) are available upon request. It should be kept in mind that our results concern emissions averaged over three summer months of 2003 rather than annual emissions. Accordingly, the emission uncertainties discussed here may not only be due to uncertainties in the corresponding annual data but also due to uncertainties in factors describing the temporal evolution of the emission in the model. Nonetheless, as gridded emission datasets are mainly intended for the use in the atmospheric models, we believe that the results of the analysis performed in this study may help in improving performance of other continental scale models.

## 5 Conclusions

We used data of tropospheric $NO_2$ column amounts derived from satellite measurements in combination with measurements of near surface ozone concentrations in order to estimate uncertainties in spatially distributed emission data used in a continental scale chemistry transport model and to elaborate less uncertain emission estimates. We employed an original inverse modeling method, which, in contrast to more common approaches, does not require explicit assumptions about the magnitudes of uncertainties in emission inventories and measurement data. Specifically, we derive the information on spatial distribution of emissions from satellite data, while the uncertainties in the input data are estimated using both ground based ozone observations and satellite data. Our inverse modeling scheme also includes a Monte Carlo experiment, which is designed to estimate uncertainties in a posteriori emissions. We used the state of-the-art chemistry transport model CHIMERE, in which the anthropogenic emissions are prescribed based on EMEP emission inventory for the year 2003 and biogenic NO emissions are calculated in the model based on well known methodologies and data. The data for tropospheric $NO_2$ columns were derived from SCIAMACHY measurements performed during the summer months of 2003 and the data for near surface ozone concentrations were obtained from the EMEP ground based monitoring network that includes more than 100 stations. The region considered in this study includes Western and Eastern Europe, a part of Northern Africa and Middle East.

Our results suggest that both the anthropogenic and biogenic emissions used in CHIMERE are rather uncertain. Specifically, the uncertainties in the anthropogenic and biogenic emissions are estimated to be about 1.7 and 7.4 (in terms of the geometric standard deviation). Such a level of emission uncertainties is found to be consistent with differences between different emission inventories. The spatial distribution of the emission uncertainties is neither homogeneous nor fully random. It is found, in particular, that the total anthropogenic and biogenic a priori emission estimates are probably mostly overestimated for Great Britain, Belgium, the Netherlands, Lithuania, Latvia, Poland, Bulgaria, Greece, and Iraq, but mostly underestimated for Spain, Italy, Switzerland, Czechia, Former Yugoslavia, Turkey, Lebanon and Iran. We also have found statistically significant differences between the a posteriori and a priori emissions for several major cities and their surroundings. In particular, the a priori emissions are probably overestimated for Baghdad (a factor of 3.1), Bucharest (1.3), Copenhagen (1.4), Enteb (1.3), London (1.2), Nizhny Novgorod (1.3), S.-Petersburg (1.8), Samara (1.3), Stockholm (1.3) and Volgograd (1.5), while the underestimation is probable for Amman ( a factor of 1.5), Beirut (3.8), Bursa (1.7), Isfahan (2.1), Istanbul (1.7), Izmir (1.7), Madrid (1.3), Tabriz (6.3), Tbilisi (3.2), Tehran (3.3) and Yerevan (3.4).

It is found that the a posteriori emissions considerably improve the agreement between the simulated and measured $NO_2$ columns. Even more importantly, a statistically significant improvement of the model performance is obtained with respect to near surface ozone concentrations. It is found that the root mean squared error defined for daily maximums of ozone concentrations is decreased for 67 percent of the stations and the correlation coefficient is increased for 63 percent of the stations. Although the obtained changes in the ozone statistics are not large, it is argued that they are consistent with the sensitivity of ozone to changes in the spatial structure of $NO_x$ emissions. This sensitivity is not strong because ozone behaviour is controlled by many other factors besides the spatial distribution of $NO_x$ emissions. Since the ozone measurements (in contrast to the data for $NO_2$ columns) have not been directly used for optimisation of the spatial structure of $NO_x$ emissions, we consider the improvement of ozone simulation as an important argument that the a posteriori emission estimates are really better than the a priori emissions. Moreover, we believe that this result is very important in view of prospects of further applications of satellite measurements in regional air pollution studies, as it shows that satellite observations can help in improving air quality models. However, this study is only one of the first steps in this promising direction.

Future work in this direction may include, in particular, studying the long-term temporal evolution of $NO_x$ emissions by using the available sets of GOME and SCIAMACHY observations since 1996. This would allow validation of the emission trends reported in available "bottom-up" emission inventories. Another promising opportunity is the inversion of emissions of hydrocarbons. Although the available observations of hydrocarbons in the lower atmosphere are rather limited, we believe that the satellite measurements of several hydrocarbon species such as formaldehyde, glyoxal, and methanol (e.g. Palmer et al., 2001; Dufour et al., 2006; Wittrock et al., 2006) can be used as indicators of emissions of other hydrocarbons. We believe also that the combination of satellite and ground based measurements can be used for obtaining more information on variations of emissions of $NO_x$ and VOC on scales from days to months.

**Acknowledgements**

I.B. Konovalov acknowledges the support of Russian Foundation for Basic Research (grant No. 05-05-64365-a), Russian Academy of Sciences (in the framework of the Programme for Basic Research "Physics of Atmosphere; Electrical Processes, Radiophysical Methods of Research"), Centre National de la Recherche Scientifique (CNRS) and the University Paris-12. M. Beekmann acknowledges support from the French LEFE-CHAT program. SCIAMACHY raw data were provided by ESA through DFD/DLR. Analysis of the SCIAMACHY data was supported by the European Union through ACCENT and by the University of Bremen. The authors acknowledge data provided by participants of the EMEP measurement network.

**References**

Beirle, S., Platt, U., Glasow, R.V., Wenig, M., Wagner T.: Estimate of nitrogen oxide emission from shipping by satellite remote sensing, Geophys. Res. Letts., 31 (18), L18102, 2004.

Beirle S., Spichtinger, N., Stohl, A., Cummins, K. L., Turner, T., Boccippio, D., Cooper, O. R., Wenig, M., Grzegorski, M., Platt, U., Wagner, T.: Estimating the $NO_x$ produced by lightning from GOME and NLDN data: a case study in the Gulf of Mexico, Atmos. Chem. Phys., 6, 1075-1089, 2006.

Bertram, T. H.; Heckel, A.; Richter, A.; Burrows, J. P.; Cohen, R. C.: Satellite measurements of daily variations in soil NOx emissions, Geophys. Res. Lett., 32(24), L24812, 2005.

Boersma, K. F., Eskes, H. J., Meijer, E. W., Kelder, H. M.: Estimates of lightning $NO_x$ production from GOME satellite observations, Atmos. Chem. Phys., 5, 2311–2331, 2005.

Bovensmann, H., Burrows, J.P., Buchwitz, M. et al.: SCIAMACHY –mission objectives and measurement modes, J. Atmos. Sci., 56, 127–150, 1999.

Bowman, K.P.: Transport of carbon monoxide from the tropics to the extratropics, J. Geophys. Res., 111, doi:10.1029/2005JD006137, 2006.

Buchwitz, M., Noël, S., Bramstedt, K., Rozanov V.V., Eisinger, M., Bovensmann, H., Tsvetkova, S., and Burrows, J.P.: Retrieval of trace gas vertical columns from SCIAMACHY/ENVISAT near-infrared nadir spectra: First preliminary results, Adv. Space Res., *34*, 809-814, 2004.

Burrows, J.P., Weber, M., Buchwitz, M., Rozanov, V.: Ladstätter-Weißenmayer, A., Richter, A., DeBeek, R., Hoogen, R., Bramstedt, K., Eichmann, K. -U., Eisinger, M., and Perner, D.: The Global Ozone Monitoring Experiment (GOME): Mission concept and first scientific results, J. Atmos. Sci., 56, 151-175, 1999.

Chipperfield, M. P.: Multiannual Simulations with a Three-Dimensional Chemical Transport Model, J. Geophys. Res., 104, 1781–1805, 1999.

Dufour, G., Boone, C. D., Rinsland, C. P., Bernath, P. F.: First space-borne measurements of methanol inside aged tropical biomass burning plumes using the ACE-FTS instrument, Atmos. Chem. Phys., 6, 3463-3470, 2006.

GENEMIS (Generation of European Emission Data for Episodes) project. EUROTRAC Annual Report 1993, Part 5, EUROTRAC International Scientific Secretariat. Garmisch-Partenkirchen, Germany, 1994.

Guenther, A., Hewitt, C. N., Erickson, D., Fall, R., Geron, C., Graedel, T., Harley, P., Klinger, L., Lerdau, M., McKay, W. A., Pierce, T., Scholes, B., Steinbrecher, R., Tallamraju, R., Taylor, J., Zimmerman, P.: A global model of natural volatile organic compound emissions. J. Geophys. Res., 100, 8873-8892, 1995.

Horowitz L. W., et al., A global simulation of tropospheric ozone and related tracers: Description and evaluation of MOZART, version 2, J. Geophys. Res., 108, 4784, doi:10.1029/2002JD002853, 2003.

Jaeglé, L., Martin, R. V., Chance, K., Steinberger, L., Kurosu, T. P., Jacob, D. J., Modi, A.I., Yoboué, V., Sigha-Nkamdjou, L., Galy-Lacaux, C.: Satellite mapping of rain-induced nitric oxide emissions from soils, J. Geophys. Res., 109 (D21310), doi:10.1029/2004JD004787, 2004.

Efron, B., Tibshirani, R. J.: An introduction to the bootstrap, New York: Chapman & Hall, 1993.

Eisinger, M., Burrows, J.P.: Tropospheric Sulfur Dioxide observed by the ERS-2 GOME instrument, Geophys. Res. Lett., 25, 4177–4180, 1998.

Enting, I.G.: Inverse problems in atmospheric constituents transport, Cambridge University Press, 2002.

Koelemeijer, R. B. A., Stammes, P., Hovenier, J. W. & de Haan, J. F., A fast method for retrieval of cloud parameters using oxygen A band measurements from the Global Ozone Monitoring Experiment, J. Geophys. Res., 106, 3475-3490, 2001.

Konovalov, I.B., Beekmann, M., Vautard, R., Burrows, J.P., Richter, A., Nüß, H., Elansky, N.: Comparison and evaluation of modelled and GOME measurement derived tropospheric $NO_2$ columns over Western and Eastern Europe, Atmos. Chem. Phys., 5, 169-190, 2005.

Konovalov, I.B., Beekmann, M., Richter, A., Burrows, J. P.: Inverse modelling of the spatial distribution of $NO_x$ emissions on a continental scale using satellite data, Atmos. Chem. Phys., 6, 1747-1770, 2006.

Lee, D. S., Köhler, I., Grobler, E., Rohrer, F., Sausen, R., Gallardo-Klenner, L., Olivier, J.J.G., Dentener, F. D.: Estimations of global NOx emissions and their uncertainties, Atmos. Environ., 31, 1735–1749, 1997.

Leue, C., Wenig, M., Wagner, T., Klimm, O., Platt, U., Jahne, B.: Quantitative analysis of NOx emissions from GOME satellite image sequences, J. Geophys. Res., 106, 5493–5505, 2001.

Martin, R.V., Jacob, D.J., Chance K., Kurosu, T., Palmer P.I., Evans, M.J.: Global inventory of nitrogen oxide emissions constrained by space-based observations of $NO_2$ columns, J. Geophys. Res., 108, 4537, doi:10.1029/2003JD003453, 2003.

Middleton, P., Stockwell, W. R., and Carter, W. P. (1990). Agregation and analysis of volatile organic compound emissions for regional modelling, Atmos. Environ., 24:1107–1133.

Mosegaard K., Tarantola A., Probabilistic Approach to Inverse Problems, in the International Handbook of Earthquake & Engineering Seismology (Part A), Academic Press, 2002, pages 237–265.

Müller, J.-F. , Stavrakou, T.: Inversion of CO and $NO_x$ emissions using the adjoint of the IMAGES model, Atmos. Chem. Phys., *5*, 1157-1186, 2005.

Palmer, P.I.; Jacob, D.J.; Chance, K.; Martin, R.V.; Spurr, R.J.D.; Kurosu, T.P.; Bey, I.; Yantosca, R.; Fiore, A., Li, Q.B.: Air mass factor formulation for spectroscopic measurements from satellites: Application to formaldehyde retrievals from the Global Ozone Monitoring Experiment, J. Geophys. Res., 106, 14539-14550, 2001.

Pétron, G., C. Granier, B. Khattatov, V. Yudin, J.-F. Lamarque, L. Emmons, J. Gille, D. P. Edwards, Monthly CO surface sources inventory based on the 2000-2001 MOPITT satellite data: Geophys. Res. Lett., 31, L21107, doi:10.1029/2004GL020560, 2004.

Press, W. H., Teukolsky, S. A., Vetterling, W. T., Flannery, B. P.: Numerical Recipes, 2nd edition, Cambridge University Press, 1992.

Richter, A., Burrows, J.P.: Tropospheric $NO_2$ from GOME measurements, Adv. Space Res., 29, 1673-1683, 2002.

Richter, A., Eyring, V., Burrows, J. P., Bovensmann, H., Lauer, A. Sierk, B., Crutzen, P. J.: Satellite Measurements of $NO_2$ from International Shipping Emissions, Geophys. Res. Lett., 31, L23110, doi:10.1029/2004GL020822, 2004.

Richter, A., Burrows, J.P., Nüß, H., Granier, C., Niemeier, U.: Increase in Tropospheric Nitrogen Dioxide Over China Observed from Space, Nature, 437, doi:10.1038/nature04092, 2005.

Simpson D., Winiwarter, W., Borjesson, G., Cinderby, S., Ferreiro, A., Guenther, A., Hewitt, C.N., Janson, R., Khalil, M.A.K., Owen, S., Pierce, T.E., Puxbaum, H., Shearer, M., Steinbrecher, S., Svensson, B.H., Tarrason, L., Oquist, M.G.: Inventorying emissions from nature in Europe, J. Geophys. Res., *104*, 8113-8152, 1999.

Stohl, A, Williams, E., Wotawa, G., Kromp-Kolb, H.: A European inventory of soil nitric oxide emissions on the photochemical formation of ozone in Europe, Atmos. Environ., 30, 3741-3755, 1996.

Tarantola, A.: Inverse problem theory; methods for data fitting and model parameter estimation, Elsevier, 1987.

Tarassón L., Jonson, J.E., Berntsen, T. K., Rypdal, K.: Study on air quality impacts of non-LTO emissions from aviation, Final report to the European Commission under contract B4-3040/2002/343093/MAR/C1, CICERO, Oslo, 2004 (http://europa.eu.int/comm/environment/air/pdf/air_quality_impacts_finalreport.pdf).

Velders G.J.M., Granier, C., Portmann, R.W., Pfeilsticker, K., Wenig, M., Wagner, T., Platt, U., Richter, A, Burrows, J.P.: Global tropospheric $NO_2$ column distributions: Comparing 3-D model calculations with GOME measurements, J. Geophys. Res., 106, 12643-12660, 2001.

Vestreng, V., Adams, M., Goodwin, J.: Inventory Review 2004. Emission data reported to CLRTAP and the NEC Directive, EMEP/EEA Joint Review report, EMEP/MSC-W Note 1, July 2004, 2004.

Visconti G., Curci, G., Redaelli, G., Grassi, B.: Synergistic Use of Satellite Data with the Global Chemistry-Transport Model GEOS-Chem: Formaldehyde Column over Europe as a proxy for Biogenic Emissions and CTM Validation using Satellite Data, in J.P Burrows and P. Borrell (eds.) Tropospheric Sounding from Space; AT2 in 2004-5, ACCENT Secretariat Urbino, 2005, Report 6.05; p. 268-272.

Wittrock F., Richter, A., Oetjen, H., Burrows, J.P., Kanakidou, M., Myriokefalitakis, S., Volkamer, R., Beirle, S., Platt, U., Wagner, T.: Simultaneous global observations of glyoxal and formaldehyde from space: Geophys. Res. Lett., 33, L16804, doi:10.1029/2006GL026310, 2006.

Yurganov, L.N., Duchatelet, P., Dzhola, A.V., Edwards, D. P., Hase, F., Kramer, I., Mahieu, E., Mellqvist, J., Notholt, J., Novelli, P. C., Rockmann, A., Scheel, H. E., Schneider, M., Schulz, A., Strandberg, A., Sussmann, R., Tanimoto, H., Velazco, V., Drummond, J.R., Gille, J.C.: Increased Northern Hemispheric carbon monoxide burden in the troposphere in 2002 and 2003 detected from the ground and from space, Atmos. Chem. Phys., 5, 563–573, 2005.

Table 1. Estimates of the mean values of the logarithmic standard deviation for uncertainties in anthropogenic ($\sigma_a$) and biogenic emissions ($\sigma_b$)

| Region | $\sigma_a$ | $\sigma_b$ |
|---|---|---|
| Western Europe | 0.5 (±0.1) | 2.0 (± 1.0) |
| Eastern Europe | 0.5 (±0.1) | 1.1 (±0.5) |
| Afro-Asian region | 0.9 (±0.2) | 1.6 (±0.7) |

Table 2. Mean uncertainty factors and the averages of the relative changes of uncertainty factors in a priori and a posteriori emissions.

| Region | $\overline{\Delta}_a$ | $\overline{\Delta}_p$ | $\overline{\left(\Delta_a - \Delta_p / \Delta_a - 1\right)}$ |
|---|---|---|---|
| Western Europe | 1.67 (1.07) | 1.53 | 0.21 |
| Eastern Europe | 1.73 (1.13) | 1.61 | 0.11 |
| Afro-Asian region | 2.15 (1.18) | 1.98 | 0.15 |

Table 3. A priori and a posteriori estimates of $NO_x$ emission rates in some major cities

| Cities | A priori emissions | A posteriori emissions |
|---|---|---|
| Alger | 3.1 (1.6) | 4.4 (1.6) |
| Aleppo | 8.1 (1.5) | 6.7 (1.3) |
| Alexandra | 2.8 (1.6) | 3.7 (1.5) |
| **Amman** | 5.8 (1.5) | 8.9 (1.4) |
| Ankara | 11.0 (1.5) | 11.2 (1.4) |
| **Baghdad** | 31.0 (1.4) | 9.9 (1.3) |
| Barcelona | 19.7 (1.3) | 19.0 (1.2) |
| **Beirut** | 1.7 (1.8) | 6.5 (2.5) |
| Belgrade | 10.6 (1.3) | 11.3 (1.3) |
| Berlin | 19.0 (1.3) | 17.6 (1.2) |
| Brussels | 47.1 (1.3) | 40.5 (1.2) |
| Budapest | 14.1 (1.3) | 12.5 (1.3) |
| **Bucharest** | 20.2 (1.4) | 15.4 (1.2) |
| **Bursa** | 11.7 (1.4) | 20.2 (1.3) |
| Chelyabinsk | 0.51 (1.7) | 0.9 (1.9) |
| **Copenhagen** | 20.0 (1.3) | 14.0 (1.2) |
| Damascus | 2.1 (1.7) | 4.1 (2.0) |
| Dnepropetrovsk | 6.7 (1.3) | 7.7 (1.3) |
| Dublin | 13.3 (1.4) | 12.5 (1.3) |
| **Enteb** | 11.9 (1.4) | 9.1 (1.3) |

| City | | |
|---|---|---|
| Hamburg | 17.4 (1.3) | 15.2 (1.2) |
| Haraj | 4.1 (1.4) | 4.0 (1.3) |
| **Isfahan** | 7.8 (1.5) | 16.2 (1.3) |
| **Istanbul** | 10.0 (1.4) | 17.0 (1.3) |
| **Izmir** | 6.7 (1.5) | 11.2 (1.5) |
| Kazan | 6.3 (1.4) | 6.8 (1.3) |
| Kharkiv | 5.5 (1.4) | 5.5 (1.3) |
| **Madrid** | 20.5 (1.3) | 26.2 (1.2) |
| Meshed | 7.4 (1.5) | 6.6 (1.3) |
| Milan | 37.3 (1.3) | 37.2 (1.2) |
| Moscow | 72.7 (1.4) | 62.2 (1.2) |
| Munich | 15.8 (1.3) | 17.3 (1.2) |
| Kiev | 9.2 (1.4) | 9.0 (1.3) |
| **London** | 59.5 (1.3) | 48.5 (1.2) |
| Minsk | 8.5 (1.3) | 10.1 (1.3) |
| **Nizhny Novgorod** | 12.5 (1.4) | 9.5 (1.3) |
| Paris | 45.0 (1.3) | 39.0 (1.2) |
| Praha | 16.9 (1.3) | 19.6 (1.3) |
| Rabat | 2.5 (1.7) | 2.6 (1.5) |
| Rome | 16.3 (1.3) | 18.2 (1.2) |
| Rostov-na-Donu | 11.3 (1.4) | 12.1 (1.3) |
| **S. -Petersburg** | 27.5 (1.4) | 14.8 (1.2) |
| **Samara** | 15.8 (1.4) | 12.1 (1.2) |
| Shiraz | 7.9 (1.4) | 9.3 (1.3) |
| Sofia | 8.8 (1.4) | 7.9 (1.3) |
| **Stockholm** | 9.3 (1.4) | 7.2 (1.3) |
| **Tabriz** | 0.7 (2.1) | 4.4 (2.1) |
| **Tbilisi** | 2.6 (1.6) | 8.4 (1.4) |
| **Tehran** | 5.7 (1.5) | 18.8 (1.6) |
| Vienna | 15.0 (1.3) | 14.7 (1.2) |
| **Volgograd** | 11.6 (1.4) | 7.7 (1.3) |
| Warsaw | 11.0 (1.3) | 9.6 (1.3) |
| Ufa | 8.8 (1.4) | 9.7 (1.3) |
| **Yerevan** | 3.0 (1.6) | 10.1 (1.6) |

The reported values represent average total emission rates for three summer months (June to August) 2003 in units of molecules$*10^{11}/(cm^2*s)$. The estimates are obtained using bi-linear interpolation of the data for 4 model's grid cells closest to the city center. The uncertainties (in terms of the geometric standard deviation) are given in the brackets. The cities for which differences between the a priori and a posteriori emissions exceed the uncertainty range are marked in bold.

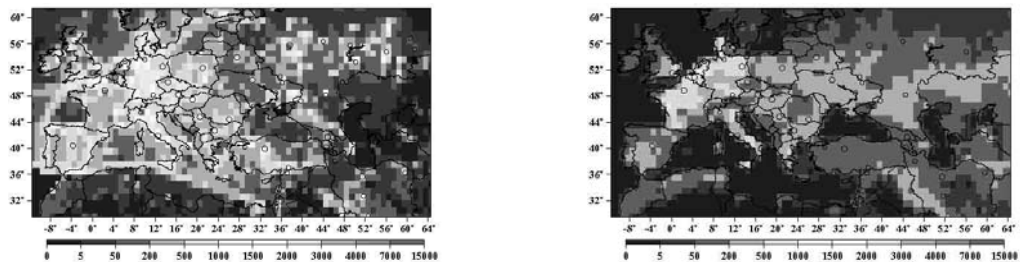

**Figure 1.** The average anthropogenic (left) and biogenic (right) emission rates (in molecules·cm$^{-2}$·s$^{-1}$·10$^8$) specified in CHIMERE for the period of June-August 2003.

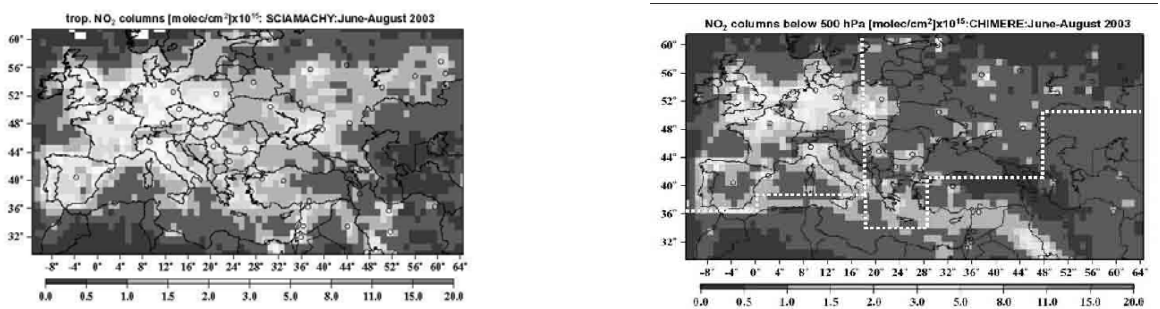

**Figure 2.** Tropospheric NO$_2$ columns derived from SCIAMACHY measurements (left) in comparison with lower tropospheric NO$_2$ columns (below 500 hPa) calculated by CHIMERE (right), the latter shows also boundaries of three regions considered in this study separately.

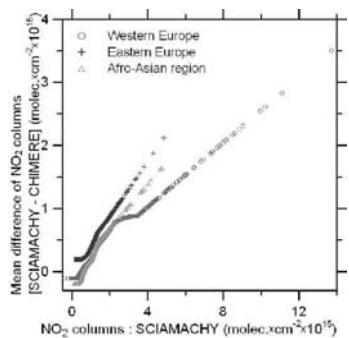

**Figure 3.** The systematic differences between the measured and modeled NO$_2$ columns estimated using the "windowing" technique as a function of magnitude of the measured columns. The window of the analysis included 100 data points.

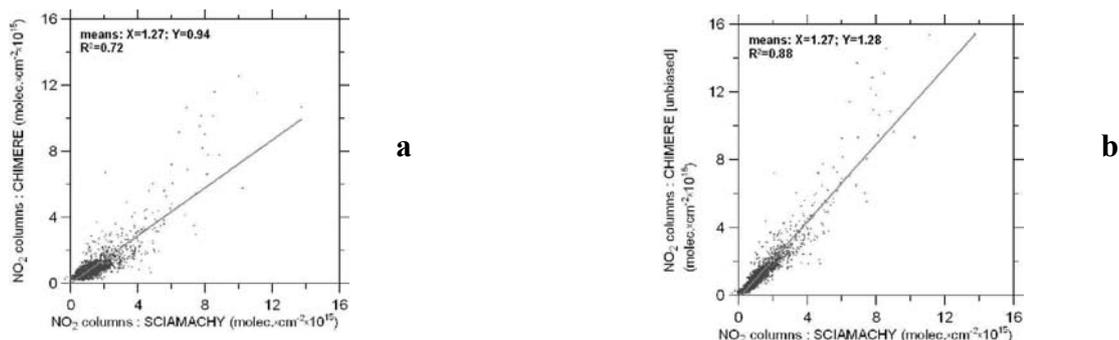

**Figure 4.** The scatter plot of the measured NO$_2$ columns versus (a) the original modelled NO$_2$ columns and (b) the modelled NO$_2$ columns with added systematic differences.

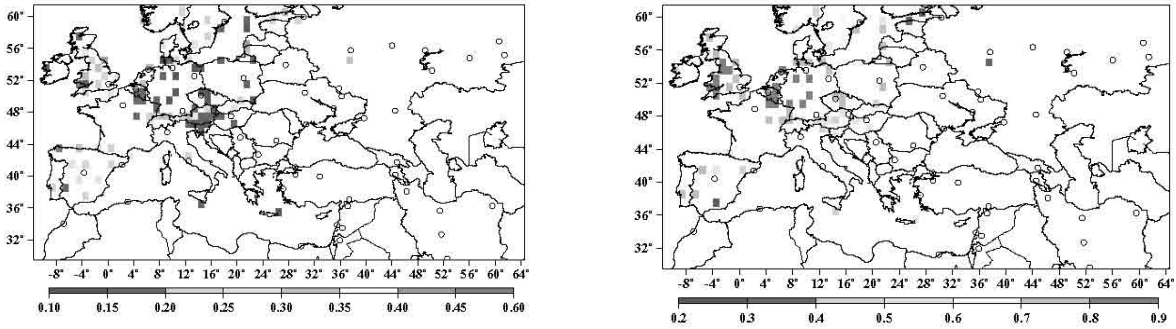

**Figure 5.** Comparison statistics for daily maximums of ozone concentrations simulated by CHIMERE and measured by ground based ozone monitors from June to August of 2003: (left) the normalized root mean squared error, (right) the correlation coefficient.

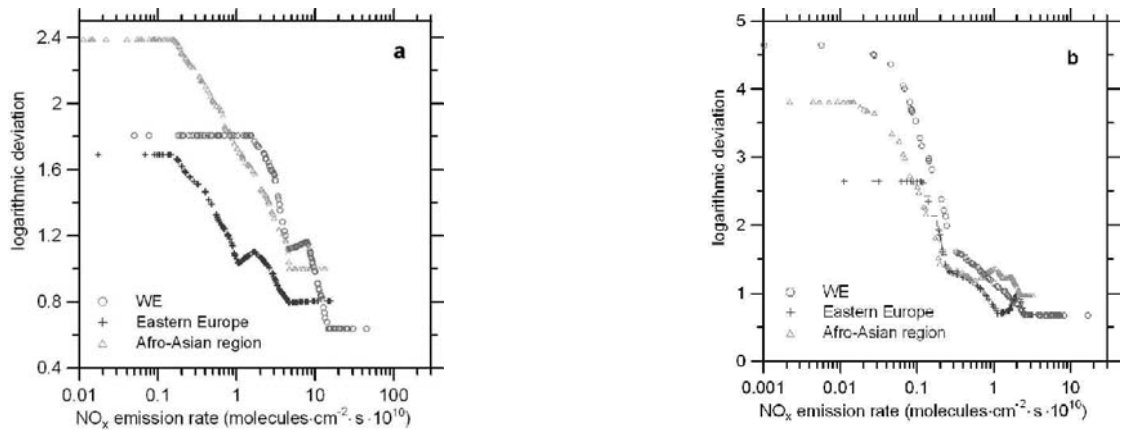

**Figure 6.** The root mean squared differences between the natural logarithms of (a) anthropogenic emission rates provided by EMEP and POET emission inventories for the year 2000 and (b) biogenic emission rates as provided by GEIA emission inventory and those calculated in CHIMERE. The differences are evaluated as a function of the magnitude of the emission rates using a windowing technique with a window comprising 100 data points. Only each fifth data point is shown.

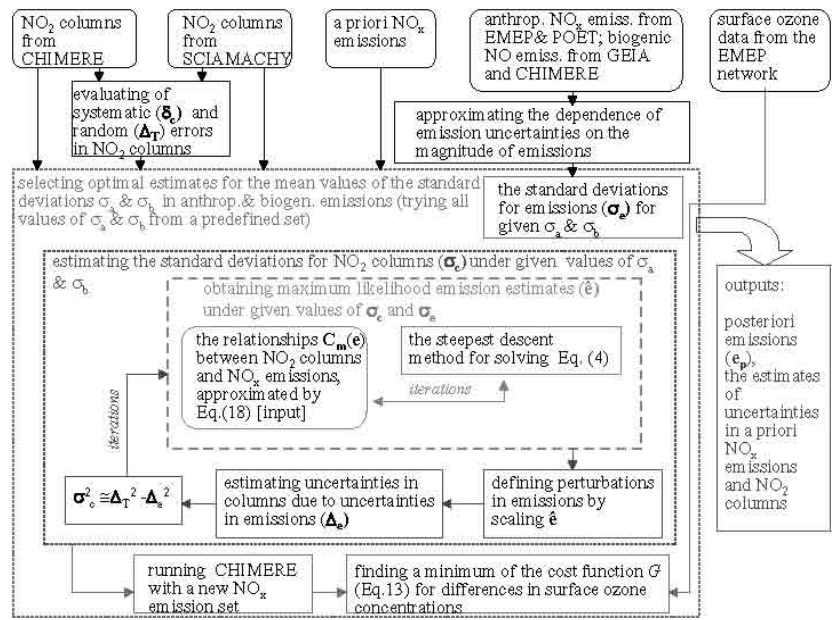

**Figure 7.** A diagram illustrating major steps of the inversion algorithm

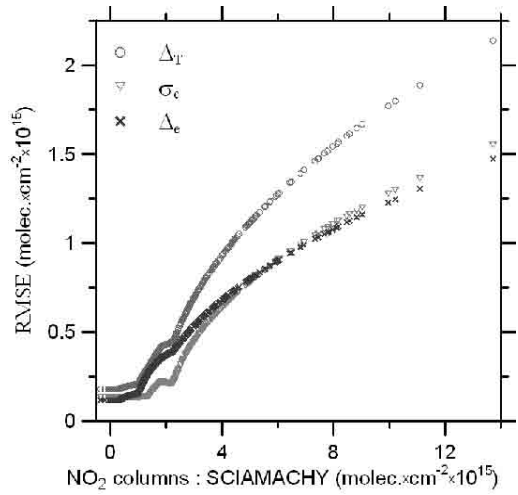

**Figure 8.** The estimates of the nonsystematic part of the total uncertainties in measured and modeled NO$_2$ columns with (circles) and without (triangles) accounting for uncertainties in a priori emissions. Also shown (crosses) is the part of uncertainties in NO$_2$ columns due to uncertainties in a priori emissions. Each point on the graph represents one grid cell of the model domain in Western Europe. The same values of $\sigma_c$ are used outside of Western Europe.

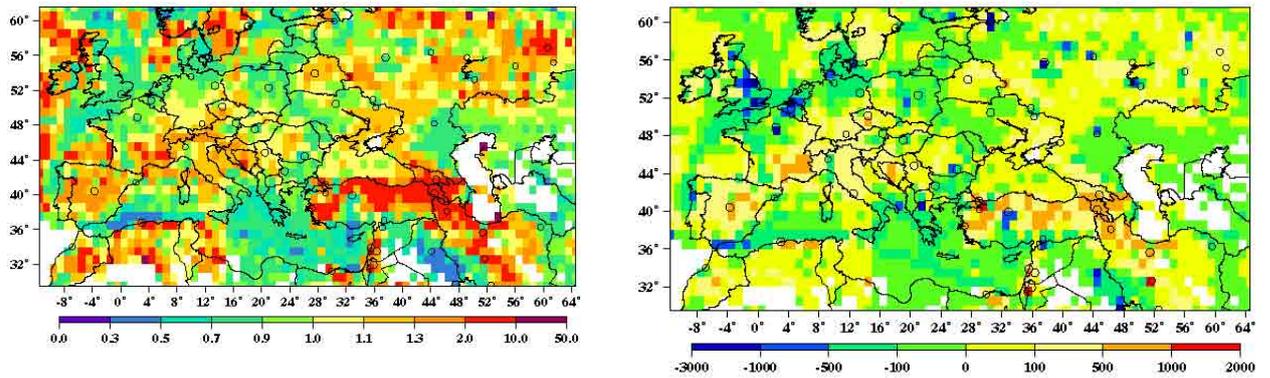

**Figure 9.** The ratios (left) and the differences (right) of the a posteriori and a priori estimates for total (anthropogenic plus biogenic) NO$_x$ emissions

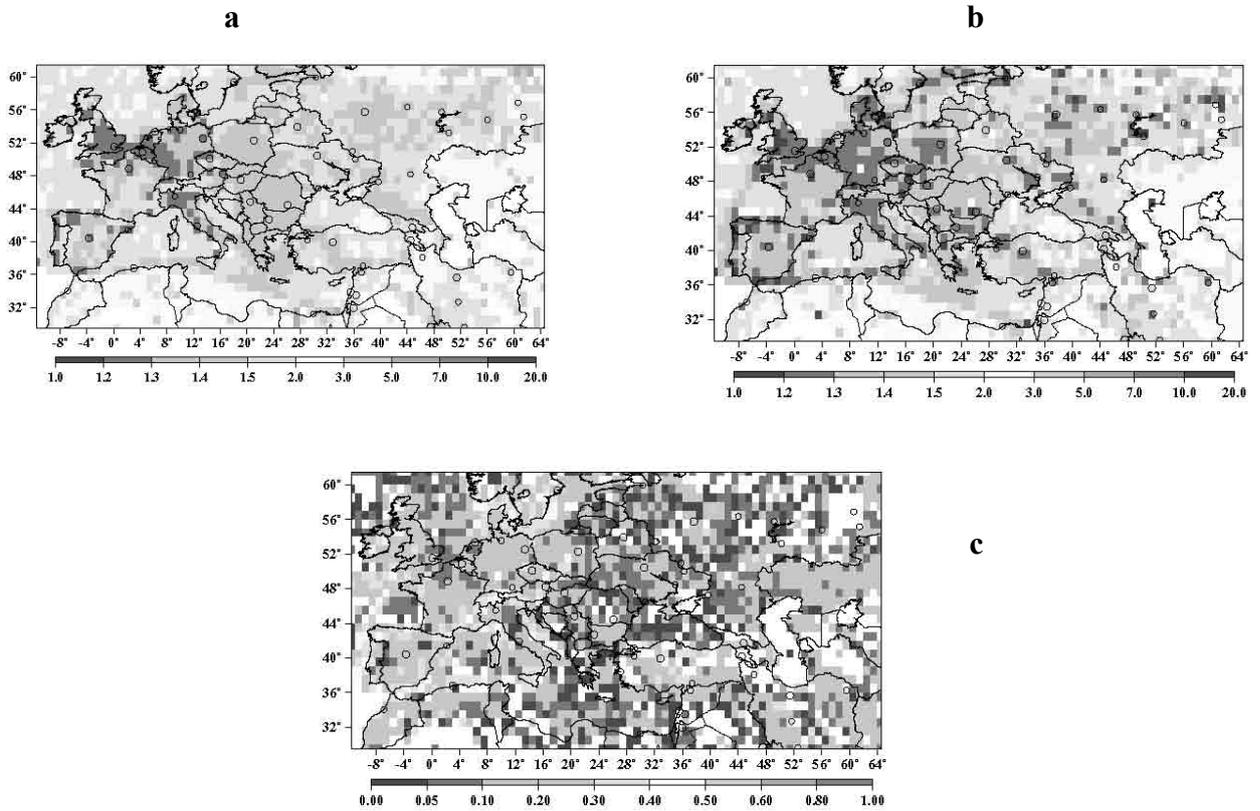

**Figure 10.** (a, b) The uncertainty factors for a priori ($\Delta_a$) and a posteriori ($\Delta_p$) emissions, respectively. (c) The relative differences [($\Delta_a-\Delta_p$)/($\Delta_a-1$)] of the uncertainties. The uncertainties are defined in terms of the 68.3 percentile. Only those grid cells are represented (not blank) in the plot "c", for which the a postreriori uncertainty is smaller than the a priori uncertainty. The plot "b" shows additionally grid cells with the statistically significant sign of the difference between the a posteriori and a priori emissions.

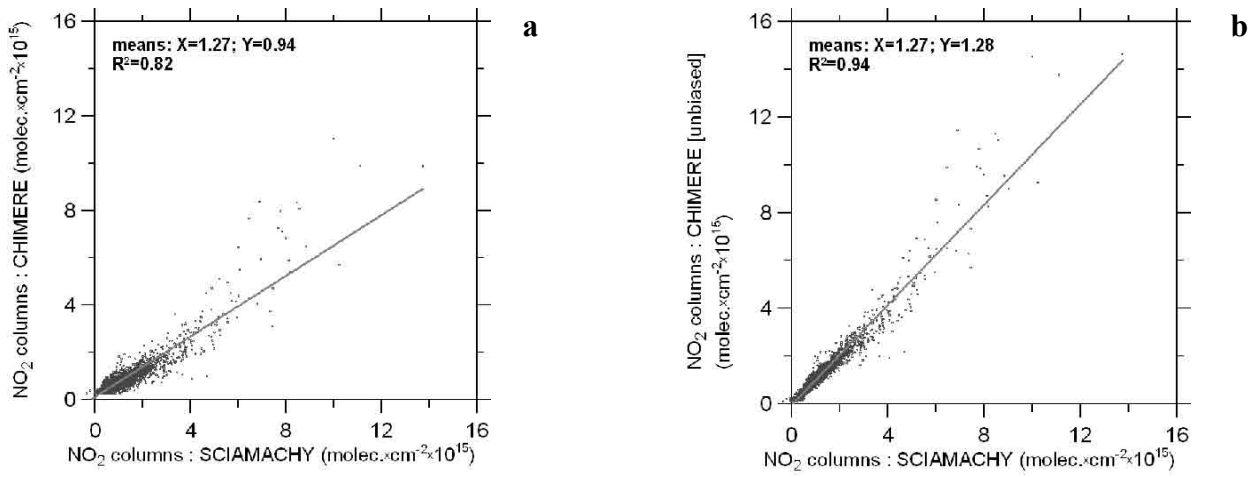

**Figure 11.** The same as in Fig.4, but with $NO_2$ columns calculated with the a posteriori emissions

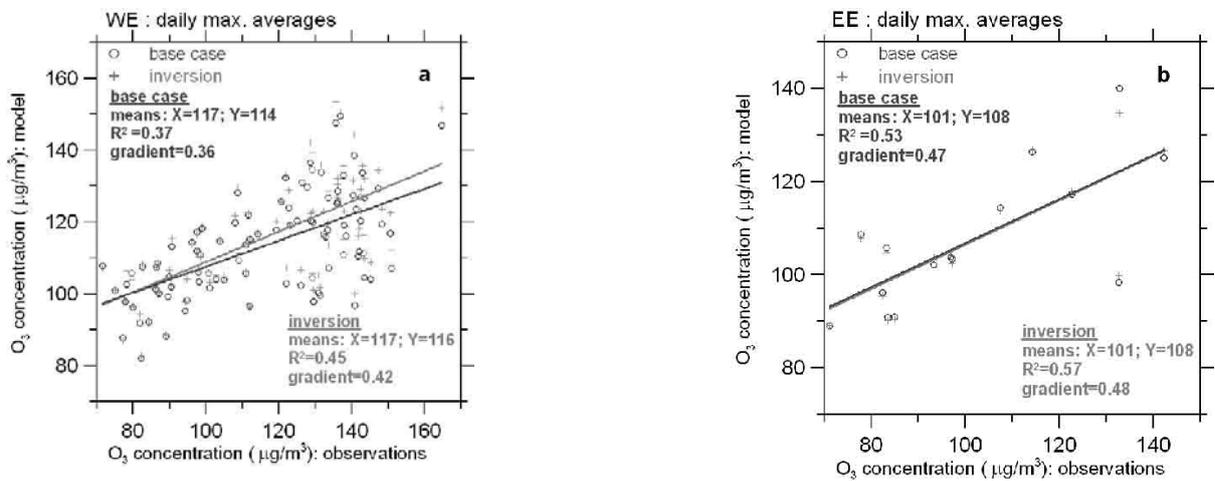

**Figure 12.** Scatter plots of the seasonally average daily maximums of ozone concentrations at the stations in (a) Western and (b) Eastern Europe

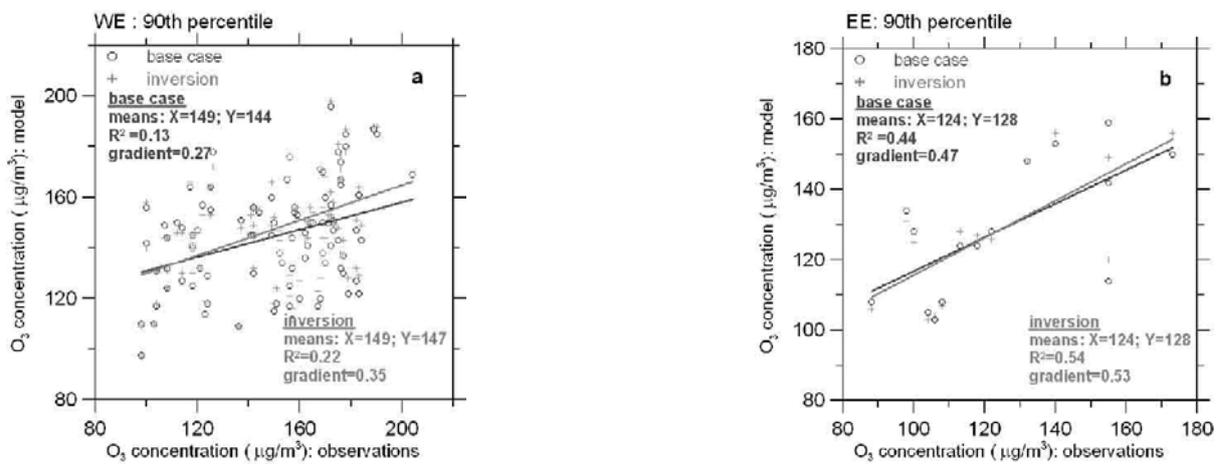

Figure 13. The same as in Fig. 12 but with the 90$^{th}$ percentiles of ozone daily maximums

22